\documentclass[aip,jmp,preprint]{revtex4-1} %reprint dà  uno stile simile a quello di pubblicazione
\usepackage[english]{babel}
\usepackage[latin1]{inputenc}
\usepackage{indentfirst}
\usepackage{amsmath, amssymb, mathrsfs, dsfont, mathtools, empheq}
\usepackage{amsthm}
\usepackage{graphicx}
\usepackage{float}
\usepackage{comment}
\usepackage{color}

\newcommand{\N}{\mathbb N}
\newcommand{\R}{\mathbb R}

\newcommand{\C}{\mathbb C}

\newcommand{\1}{\mathds 1}

\newcommand{\dd}{\mathrm d}
\newcommand{\Card}{\textit{Card}}

\DeclareMathOperator*{\argmax}{arg\,max}

\DeclarePairedDelimiter{\floor}{\lfloor}{\rfloor}
\newcommand{\dsum}{\displaystyle\sum}

\newcommand{\tprod}{\textstyle\prod}
\newcommand{\DD}{\mathscr D}
\newcommand{\MM}{\mathscr M}
\newcommand{\w}{\mathbf w}
\newcommand{\x}{\mathbf x}

\newcommand{\hm}{h^{\textup{\tiny{(m)}}}}
\newcommand{\hd}{h^{\textup{\tiny{(d)}}}}
\newcommand{\Jm}{J^{\textup{\tiny{(m)}}}}
\newcommand{\Jd}{J^{\textup{\tiny{(d)}}}}
\newcommand{\Jmd}{J^{\textup{\tiny{(md)}}}}
\newcommand{\Him}{H^{\textup{\tiny{IMD}}}}
\newcommand{\Zim}{Z^{\textup{\tiny{IMD}}}}
\newcommand{\muim}{\mu^{\textup{\tiny{IMD}}}}
\newcommand{\pim}{p^{\textup{\tiny{IMD}}}}
\newcommand{\mim}{m^{\textup{\tiny{IMD}}}}
\newcommand{\vpim}{\widetilde{p}\,}%^{\textup{\tiny{(im)}}}}
\newcommand{\Hcl}{H^{\textup{\tiny{MD}}}}
\newcommand{\Zcl}{Z^{\textup{\tiny{MD}}}}
\newcommand{\mucl}{\mu^{\textup{\tiny{MD}}}}
\newcommand{\pcl}{p^{\textup{\tiny{MD}}}}
\newcommand{\mcl}{m^{\textup{\tiny{MD}}}}
\newcommand{\Hising}{H^{\textup{\tiny{ISING}}}}
\newcommand{\Zising}{Z^{\textup{\tiny{ISING}}}}

\theoremstyle{plain}
\newtheorem{thm}{Theorem}
\newtheorem{lem}{Lemma}
\newtheorem{prop}{Proposition}
\newtheorem{cor}{Corollary}

\theoremstyle{definition}
\newtheorem{df}{Definition}
\newtheorem{rk}{Remark}

\begin{document}
\preprint{AIP/123-QED}

\title{A mean-field monomer-dimer model with attractive interaction. The exact solution.}

\author{\vspace{1cm}D. Alberici} \email{diego.alberici2@unibo.it}
\author{P. Contucci} \email{pierluigi.contucci@unibo.it}
\author{E. Mingione} \email{emanuele.mingione2@unibo.it}
\affiliation{Department of Mathematics, University of Bologna}

\date{\today}

\begin{abstract}
A mean-field monomer-dimer model which includes an attractive interaction
among both monomers and dimers is introduced and its exact solution rigorously derived.
The Heilmann-Lieb method for the pure hard-core interacting case is used
to compute upper and lower bounds for the pressure. The bounds are shown to
coincide in the thermodynamic limit for a suitable choice of the monomer density
$m$. The consistency equation characterising $m$ is studied in the phase space
$(h,J)$, where $h$ tunes the monomer potential and $J$ the attractive potential.
The critical point and exponents are computed and show that
the model is in the mean-field ferromagnetic universality class.
\end{abstract}


\maketitle

\section{Introduction and results}
Each way to fully cover the vertices of a finite graph $G$ by non-overlapping
dimers (molecules which occupy two adjacent vertices) and monomers
(molecules which occupy a single vertex) is called a monomer-dimer configuration.
Associating to each of those configurations a probability proportional to the
product of a factor $w>0$ for each dimer and a factor $x>0$ for each monomer
defines a monomer-dimer model with pure hard-core interaction.

Those models were proposed to investigate the properties of diatomic oxygen molecules
deposited on tungsten\cite{R} or to study liquid mixtures in which the molecules are unequal in size\cite{FR}.
The hard-core interaction accounts for the contact repulsion generated by the Pauli principle.
In order to account also for the attractive component of the Van der Waals potential among monomers and dimers,
one may consider an attractive interaction\cite{RM, Ch, Chcambr} among particles occupying neighbouring sites
(as it was previously done for single atoms\cite{F, Pe}).

More recently monomer-dimer models on diluted network have attracted a considerable attention\cite{mz,ac}
and they have been applied, with the addition of a ferromagnetic imitative interaction,
also in social sciences\cite{imm}.

The partition function describing a general system of interacting monomers and dimers
can be written as
\begin{equation}\label{main}
Z_G \; = \; \sum_{D\in \DD_G}x^Mw^{|D|}z_1^{I_m}z_2^{I_d}z_3^{I_{md}} \; ,
\end{equation}
where $z_1,z_2,z_3>0$ tune the interaction among particles and for a given dimer configuration $D$, $M$ is the
corresponding number of monomers, $I_m$ the
number of neighbouring monomers, $I_d$ the number of neighbouring dimers, $I_{md}$ the number of
neighbouring molecules of different type.

In this paper we investigate a system where the attraction among monomers and among dimers is stronger than
the attraction among molecules of different type, that is $z_1z_2\geq z_3^2$.
And precisely we study the mean-field case, i.e. the model on the complete graph where each of the $N$ sites
is connected with all the others and the particle system is permutation invariant.
Considering the relation $2|D|+M=N$ induced by the hard-core interaction among particles,
we may study without loss of generality a reduced model given by the parametrisation $x=e^h$, $w=1/N$,
$z_1=z_2=e^{J/N}$, $z_3=1$.
We prove that, at large volumes, the model turns out to be described by the monomer density $m(h,J)$,
i.e. the expectation value, with respect to the probability measure introduced by (\ref{main}),
of the fraction of sites occupied by monomers.

For pure hard-core interactions, i.e. $J=0$, Heilmann and Lieb \cite{HL, HLprl} proved the absence
of phase transitions for both regular lattices and in the mean-field case (complete graph) treated here.
Using the relation between the partition function and the Hermite polynomials, we compute here the thermodynamic limit
of the free energy in the pure hard-core case and use it to solve the attractive case by means of
a one-dimensional variational principle in the monomer density.
For a suitable smooth, monotonic, function $g$ mapping $\R$ into the interval $(0,1)$,
we find that $m(h,J)$ can be identified among the solutions (at most three) of the consistency equation
\begin{equation}\label{mdce}
m=g((2m-1)J+h)
\end{equation}
characterising the entire phase space of the model.
In particular it turns out that $m$ has, in the $(h,J)$ plane, a jump discontinuity on a curve $h=\gamma(J)$.
The curve $\gamma$, implicitly defined, stems at
\begin{equation}
(h_c,J_c) \; = \; \left( \frac{1}{2}\,\log(2\sqrt{2}-2)-\frac{1}{4}\,,\,\frac{1}{4\,(3-2\sqrt{2})} \right) \,,
\end{equation}
is smooth outside the critical point $(h_c,J_c)$ and at least differentiable approaching it,
moreover is has an asymptote at $h=-1/2$ for large values of $J$.
The order parameter $m(h,J)$ is characterised in a neighbourhood of the critical point by the mean-field
theory critical exponents: $\boldsymbol{\beta}=1/2$ along the direction of $\gamma$, and
$\boldsymbol{\delta}=3$ along any other direction of the plane $(h,J)$.

The paper is organised as follows: in Section \ref{sec: m-d classic} we introduce and solve the model without attraction
following the methods of Heilmann and Lieb. In Section \ref{sec: m-d imitative} we introduce the model with attractive interaction and we show how to control the thermodynamic limit of the free energy by means of a one
dimensional variational problem. Section \ref{sec: analysis} presents the study of the consistency equation (\ref{mdce}) in the
$(h,J)$ plane, contains the study of the implicit equation for the curve $\gamma$ and the computation of critical exponents of
the model. The Appendix contains supplementary material of elementary type that makes the paper self-contained.
\section{Monomer-dimer model}\label{sec: m-d classic}
Let $G=(V,E)$ be a finite simple graph with vertex set $V$ and edge set $E\subseteq\{uv\equiv\{u,v\}\,|\,u\neq v\in V\}\,$.

\begin{df} \label{def: space of states}
A \textit{dimer configuration} $D$ on the graph $G$ is a set of pairwise non-incident edges (called \textit{dimers}):
\[ D\subseteq E \quad\text{and}\quad \big(uv\in D\ \Rightarrow\ uw\notin D\ \forall w\neq v\big) \,.\]
Given $D$, the associated \textit{monomer configuration} is the set of dimer-free vertices (called \textit{monomers}):
\[ \MM(D):=\MM_G(D):=\{u\in V\,|\,uv\notin D\ \forall v\in V \}\, .\]
Notice that $|\MM(D)|+2\,|D|=|V|\,$.
\end{df}

\begin{df} \label{def: classical m-d}
Let $\DD_G$ be the set of all possible dimer configurations on the graph $G\,$.
The \textit{monomer-dimer model} on $G$ is obtained by assigning a monomer weight $x_v>0$ to each vertex $v\in V$ and a dimer weight $w_e>0$ to each edge $e\in E$ and considering the following probability measure on the set $\DD_G$:
\[ \mucl_{G}(D)=\frac{1}{\Zcl_G(\x,\w)}\;\tprod_{e\,\in D}w_e\,\tprod_{v\,\in\MM(D)}x_v \quad\forall\,D\in\DD_G \,.\]
The normalising factor, called \textit{partition function} of the model, is
\begin{equation}\label{eq: Z classic}
\Zcl_G(\x,\w):=\sum_{D\in\DD_G}\,\tprod_{e\,\in D}w_e\,\tprod_{v\,\in\MM(D)}x_v
\end{equation}
Its natural logarithm $\log \Zcl_G$ is called \textit{pressure}.
%The quantity $\frac{1}{|V|}\,x\,\frac{\partial}{\partial x}\log \Zcl_G$ is called \textit{monomer density}.\\
%The expected value with respect to the measure $\mucl_{G,x}$ is denoted by $\bracket{\,\cdot\,}_{G,x}\,$, namely for any function $f$ of the dimer configuration
%$$\bracket{f}_{G,x}:=\sum_{D\in\DD_G}f(D)\,\mucl_{G,x}(D)\, .$$
\end{df}

\begin{rk} \label{rk: scale parameters}
If uniform dimer (resp. monomer) weights are considered, i.e. $w_e\equiv w\;\forall e\in  E$ (resp. $x_v\equiv x\;\forall v\in V$), then it's possible to keep $w=w_0$ (resp. $x=x_0$) fixed and study only the dependence of the model on $\x$ (resp. $\w$) without loss of generality. Indeed, using the relation $|\MM(D)|+2\,|D|=|V|\,$, it's easy to check that
\begin{gather}
\Zcl_G(\x,w) \,=\, (w/w_0)^{|V|/2}\;\Zcl_G\big(\frac{\x}{(w/w_0)^{1/2}}\,,\,w_0\big)\;; \label{eq: scale parameters w0}\\
\Zcl_G(x,\w) \,=\, (x/x_0)^{|V|}\;\Zcl_G\big(x_0\,,\frac{\w}{(x/x_0)^2}\big)\;. \label{eq: scale parameters x0}
\end{gather}
\end{rk}

\begin{rk} \label{rk: monomer density}
With uniform monomer weights, a direct computation shows that the \textit{monomer density}, i.e. the expected fraction of monomers on the graph, is related to the derivative of the pressure w.r.t. $x\,$:
\[ \mcl_G := \sum_{D\in\DD_G}\frac{|\MM(D)|}{|V|}\; \mucl_{G}(D) \,=\, x\,\frac{\partial}{\partial x}\frac{\log \Zcl_G}{|V|} \,.\] %(x,\w)
\end{rk}

\begin{rk} \label{rk: pressure bounds}
With bounded monomer and dimer weights $\underline x\leq x_v\leq\overline x$, $w_e\leq\overline w$, the following bounds for the pressure hold:
\[\log\underline x \,\leq\, \frac{\log \Zcl_G(\x,\w)}{|V|} \,\leq\, \log\overline x + \frac{|E|}{|V|}\,\log\big(1+\frac{\overline w}{\overline x^2}\,\big)\,.\]
\end{rk}

\proof
The lower bound is obtained from (\ref{eq: Z classic}) considering only the empty dimer configuration (i.e. a monomer on each vertex of the graph):
\[ \Zcl_G \,\geq\, \prod_{v\in V}x_v \,\geq\, \underline x^{|V|}\,.\] %(\x,\w)
The upper bound is obtained from (\ref{eq: Z classic}) using the fact that any dimer configuration made of $d$ dimers is a (particular) set of $d$ edges:
\[\begin{split}
\Zcl_G \,&\leq\, \sum_{d=0}^{|E|}\Card\{D\in\DD_G\,,\,|D|=d\}\;\overline w^d\;\overline x^{|V|-2d} \,\leq\,
\sum_{d=0}^{|E|}{|E|\choose d}\,\overline w^d\;\overline x^{|V|-2d} \,=\\
&=\,\overline x^{|V|}\,(1+\overline w\,\overline x^{-2})^{|E|}\;.\qedhere
\end{split}\] %(\x,\w)
\endproof

The following recursion for the partition function, due to Heilmann and Lieb \cite{HL}, is a fundamental property of the monomer-dimer model.
\begin{prop} \label{prop: HL recursion}
Given a vertex $o$ and its neighbours $v$, it holds
\[ \Zcl_G(\x,\w) \,=\, x_o\,\Zcl_{G-o}(\x',\w') \,+\, \sum_{v\sim o}\,w_{ov}\,\Zcl_{G-o-v}(\x'',\w'')\;,\]
where $\x',\w',\x'',\w''$ are the weights vectors conveniently restricted to the involved subgraphs.
\end{prop}

\proof
The dimer configurations on $G$ having a monomer on the vertex $o$ coincide with the dimer configurations on $G-o$. Instead the dimer configurations on $G$ having a dimer on the edge $ov$ are in one-to-one correspondence with the dimer configurations on $G-o-v$.
Therefore $\phantom\qedhere$
\[\begin{split}
\Zcl_G \,&=\,
\dsum_{D\in\DD_G} \tprod_{e\,\in D}w_e\,\tprod_{v\,\in\MM_G(D)}x_v \\
&= \dsum_{D\in\DD_G,\atop o\in\MM_G(D)}\!\!\!\! \tprod_{e\,\in D}w_e\,\tprod_{v\,\in\MM_G(D)}x_v \,+\,
\dsum_{v\sim o} \dsum_{D\in\DD_G,\atop ov\in D}\!\! \tprod_{e\,\in D}w_e\,\tprod_{v\,\in\MM_G(D)}x_v \\
&=\, x_o\!\!\dsum_{D\in\DD_{G-o}}\!\!\! \tprod_{e\,\in D}w_e\,\tprod_{v\,\in\MM_{G-o}(D)}x_v \,+\,
\dsum_{v\sim o}\, w_{ov}\!\!\!\!\!\!\dsum_{\ \ D\in\DD_{G-o-v}}\!\!\!\!\!\!\!\!\tprod_{e\,\in D}w_e\,\tprod_{v\,\in\MM_{G-o-v}(D)}x_v \\
&\,=\,x_o\,\Zcl_{G-o} \,+\, \dsum_{v\sim o}w_{ov}\,\Zcl_{G-o-v} \;. \hspace{217pt}\qed
\end{split}\]
\endproof
\subsection{The monomer-dimer model on the complete graph}
Let $K_N=(V_N,E_N)$ be the complete graph over $N$ vertices, that is $V_N=\{1,\dots,N\},\ E_N=\{uv\,|\,u,v\in V_N,\,u<v\}$. Notice $|E_N|=N(N-1)/2$.\\
We work with uniform weights and we want $\log \Zcl_{K_N} \,=\, \mathcal{O}(N)$. For this purpose, observing remark \ref{rk: pressure bounds}, we have to choose $x,\,w$ such that $w/x^2=\mathcal{O}(1/N)$. %As by remark \ref{rk: scale parameters} the partition function depends only on the ratio $w/x^2$, we can study without loss of generality
By remark \ref{rk: scale parameters} we can fix without loss of generality $w=1/N$ and study
\begin{equation} \label{eq: Z classic complete graph}
\Zcl_N(x) \,:=\, \Zcl_{K_N}\big(x\,,\frac{1}{N}\,\big)\,,
\end{equation}
indeed choosing $w_0=1/N$ in (\ref{eq: scale parameters w0}) it's easy to check that %$\Zcl_{K_N}(x,w/N)=w^{N/2}\,\Zcl_N(x/w^{1/2})$ for any $w>0$, and more generally
$\Zcl_{K_N}(x,w)=(wN)^{N/2}\,\Zcl_N(c^{-1/2})$ whenever $w/x^2=c/N$.
Observe that the bounds of remark \ref{rk: pressure bounds} become
\[\log x \,\leq\, \frac{\log \Zcl_N(x)}{N} \,\leq\, \log x + \frac{N-1}{2}\,\log\big(1+\frac{1}{N x^2}\,\big)
\,\leq\, \log x + \frac{1}{2\,x^2} \,.\]

On the complete graph it is possible to compute explicitly the partition function and it turns out to be related to the Hermite polynomials. We will give two proofs: the first one due to Heilmann and Lieb \cite{HL} is based on a recurrence relation and applies also to other graphs, the second one is based on a simple combinatorial argument. %\textit{(cit. ? Wikipedia)}.

\begin{thm}\label{thm: mean field classic}
The partition function of the monomer-dimer model on the complete graph $K_N$ is
\[ \Zcl_N(x) \,=\, \big(\frac{i}{\sqrt{N}}\,\big)^N\;H_N\big(\!-i\,x\sqrt{N}\,\big) \;,\]
where $H_N$ denotes the $N^{th}\!$ probabilistic Hermite polynomial.
\end{thm}

\proof[First proof]
Use the Heilmann-Lieb recursion of proposition \ref{prop: HL recursion} with $o=N$
\[ \Zcl_{K_N}(x,1) \,=\, x\,\Zcl_{K_N-N}(x,1) \,+\, \sum_{v=1}^{N-1} \Zcl_{K_N-N-v}(x,1) \,;\]
then observe that for any $u,\,v\in V_N$ the graphs $K_N-u$, $K_N-u-v$ are isomorphic to $K_{N-1}$, $K_{N-2}$ respectively and complete with the initial conditions:
\begin{equation}\label{eq: complete graph recursion}
\left\{\begin{array}{l}
\Zcl_{K_N}(x,1) \,=\, x\,\Zcl_{K_{N-1}}(x,1) \,+\, (N-1)\,\Zcl_{K_{N-2}}(x,1) \\[4pt]
\Zcl_{K_1}(x,1)\,=\,x\,,\quad \Zcl_{K_0}(x,1)\,=\,1
\end{array}\right.\,.
\end{equation}
Now the probabilistic Hermite polynomials are the solution of the following problem \cite{AS}
\begin{equation}
\left\{\begin{array}{l}
H_N(x) \,=\, x\,H_{N-1}(x) \,-\, (N-1)\,H_{N-2}(x)\\[4pt]
H_1(x)\,=\,x\,,\quad H_0(x)\,=\,1
\end{array}\right.\,;
\end{equation}
hence it's easy to check that the polynomials $\big\{i^N\,H_N(-i\,x)\big\}_{N\in\N}$ are the solution of problem (\ref{eq: complete graph recursion}).
Therefore $\Zcl_{K_N}(x,1) = i^N H_N(-i\,x)\,$.
Conclude using definition (\ref{eq: Z classic complete graph}) and identity (\ref{eq: scale parameters x0}) with $w=1/N$, $w_0=1$.
%\[ \Zcl_N(x) \,=\, \Zcl_{K_N}\big(x\,,\frac{1}{N}\,\big) \,=\,
%\big(\frac{1}{N}\big)^{N/2} \Zcl_{K_N}\big(x\sqrt{N}\,,1\big) \,=\,  \big(\frac{i}{\sqrt{N}}\,\big)^{\,N}\,H_N\big(-i\,x\sqrt{N}\,\big) \;.\qedhere\]
\endproof

\proof[Second proof]
In general the partition function admits the following expansion
\[ \Zcl_N(x) \,=\, \Zcl_{K_N}\big(x\,,\frac{1}{N}\big) \,=\, \sum_{d=0}^{\floor{N/2}}\,c_N(d)\;N^{-d}\;x^{N-2d} \;, \]
where $c_N(d) = \Card\{D\in\DD_{K_N}\,,\;|D|=d\}\,$.
On the complete graph these coefficients can be computed with a combinatorial argument.
Any dimer configuration $D$ on $K_N$ composed of $d$ dimers can be built by the following iterative procedure:
\begin{itemize}
\item choose two different vertices $u$ and $v$ in $V^{(s)}$ (it can be done in ${|V^{(s)}|\choose 2}$ different ways) and marry them by a dimer setting $D^{(s)}:=D^{(s-1)}\cup\,uv\,$,
\item now exclude the two married vertices setting $V^{(s+1)}:=V^{(s)}\smallsetminus\{u,\,v\}$ ;
\end{itemize}
repeat for $s=1,\,\dots,\,d$, with initial sets $V^{(1)}:=V_N$, $D^{(0)}:=\varnothing$ and finally $D:=D^{(d)}$.\\
Thus the number of possible dimer configurations with $d$ dimers on the complete graph is
\begin{equation}
c_N(d) \,=\, {N\choose2}\,{N-2\choose2}\,\dots\,{N-2(d-1)\choose2}\;\big/\;d! \,=\, \frac{N!}{d!\,(N-2d)!}\;2^{-d} \;,
\end{equation}
where in the first combinatorial computation one divides by $d!$ as not interested in the order of the $d$ dimers.
Substitute these coefficients in the expansion of the partition function:
\begin{equation} \label{eq: complete graph expansion}
\Zcl_N(x) = \sum_{d=0}^{\floor{N/2}}\,\frac{N!}{d!\,(N-2d)!}\ (2 N)^{-d}\; x^{N-2d} \ .
\end{equation}
Now the probabilistic Hermite polynomials admit the following expansion \cite{AS}:
\begin{equation} \label{eq: hermite expansion}
H_N(x) \,=\, \sum_{d=0}^{\floor{N/2}}(-1)^d\,\frac{N!}{d!\,(N-2d)!}\ 2^{-d}\; x^{N-2d} \;.
\end{equation}
Comparing (\ref{eq: complete graph expansion}) and (\ref{eq: hermite expansion}) it's easy to conclude.
%\[ \Zcl_N(x) \,=\, \big(\frac{i}{\sqrt{N}}\,\big)^N\;H_N\big(\!-i\,x\sqrt{N}\,\big) \;. \qedhere\]
\endproof

Using theorem \ref{thm: mean field classic}, and precisely formula (\ref{eq: complete graph expansion}), we explicitly compute the pressure in the limit $N\to\infty$.

\begin{prop} \label{prop: mean field classic limit}
The pressure per particle on the complete graph admits thermodynamic limit:
\[ \forall\,x\!>0\quad\exists\;\lim_{N\to\infty}\frac{\log \Zcl_N(x)}{N} \,=\,\pcl(x) \]
and $\pcl$ is a analytic function of $x>0$, precisely:
% since log and sqrt are analytic outside 0
\begin{gather}
\pcl(x) \,=\, f(x)\,\big(1-\log f(x)-\log 2\big) + g(x)\,\big(1-\log g(x)+\log x\big) - 1 \;, \label{eq: p classic mean field 1} \\[2pt]
f(x) \,=\, \frac{1}{4}\;\big(2+x^2-\sqrt{x^4+4x^2}\,\big) \;\in\,]0,\frac{1}{2}[ \;, \label{eq: f(x)}\\
g(x) \,=\, 1-2\,f(x) \,=\, \frac{1}{2}\,(\sqrt{x^4+4x^2}-x^2) \;\in\,]0,1[ \;. \label{eq: g(x)}
\end{gather}
\end{prop}

\proof It is convenient to set for $d=0,\dots,\floor{N/2}$
\[ a_N(d,x) := \frac{N!}{d!\,(N-2d)!}\ (2N)^{-d}\; x^{N-2d} \quad,\quad
M_N(x) := \max_{d=0\dots\floor{N/2}}a_N(d,x) \;.\]
By formula (\ref{eq: complete graph expansion}) the explicit expansion of the partition function is
\[\Zcl_N(x)=\sum_{d=0}^{\floor{N/2}}a_N(d,x) \;,\\[-4pt]\]
hence $M_N(x) \leq \Zcl_N(x) \leq (\frac{N}{2}+1)\,M_N(x)$
and taking the $\log$ and dividing by $N$ one obtains
\[\frac{\log M_N(x)}{N} \,\leq\, \frac{\log \Zcl_N(x)}{N} \,\leq\, \underbrace{\frac{\log\big(\frac{N}{2}+1\big)}{N}}_{\xrightarrow[N\to\infty]{}\,0} \,+\, \frac{\log M_N(x)}{N} \;.\\[-2pt]\]
Therefore if one proves that $(\log M_N)/N\rightarrow l\,$ as $N\to\infty$, it will follow that also $(\log \Zcl_N)/N\rightarrow l\,$ as $N\to\infty$. Let's study the asymptotic behaviour of $(\log M_N)/N\,$.\\[4pt]
\textit{I.} The first step is to understand which is the maximum term of each sum, studying the trend of $a_N(d,x)$ as a function of $d\in\{\,0,\,\dots,\floor{N/2}\}\,$.\\
Simplifying factorials and powers and isolating $d$ and $d^2$, one finds
\[ a_N(d,x) \,\leq\, a_N(d+1,x)\ \ \Longleftrightarrow\ \ 4\,d^2 - 2\,(2N-1+Nx^2)\,d + N\,(N-1-2x^2) \geq 0 \quad(\diamond) \]
Solve this second degree inequality in $d$, finding $d\leq d_-(N,x)$ or $d\geq d_+(N,x)$.
%\[\begin{split}
%&d \,\leq\, \frac{1}{4}\,(2N-1+N x^2) - \frac{1}{8}\,\sqrt{\Delta(N,x)}\ =: d_-(N,x)\ \ \ \text{or}\\[2pt]
%&d \,\geq\, \frac{1}{4}\,(2N-1+N x^2) + \frac{1}{8}\,\sqrt{\Delta(N,x)}\ =: d_+(N,x) \;,
%\end{split}\]
%with
%\[\Delta(N,x) = 4\,\big[(x^4+4x^2)\,N^2 + 6x^2\,N + 1\big] > 0 \,.\]
For $N\rightarrow\infty$ one may estimate
\[ d_{\pm}(N,x) \,=\, f_{\pm}(x)\,N + \mathcal{O}(\sqrt{N}\,)\,, \quad\text{with}\quad f_{\pm}(x) \,:=\, \frac{1}{4}\,\big(2+x^2\pm \sqrt{x^4+4x^2}\,\big) \,.\]
Observe that $f_+(x)>1/2\,$ while $f_-(x)<1/2\,$, hence for $N$ sufficiently large $d_+(N,x)>N/2$ while $d_-(N,x)<N/2$.
Therefore the inequality $(\diamond)$ with $d\leq N/2$ is equivalent to $d\leq d_-(N,x)\,$.
To resume, for $N$ sufficiently large
\[ a_N(d,x) \,\leq\, a_N(d+1,x)\ \ \Longleftrightarrow\ \ d \,\leq\, d_-(N,x) \,=\, f_-(x)\,N + \mathcal{O}(\sqrt{N}\,) \;.\]
\textit{II.} Now knowing that the maximum term of the sum is the one with index $d=d_{\max}=\floor{d_-(N,x)}+1$, compute
\[\begin{split}
M_N(x) \,&=\, \max_{d=0\dots\floor{N/2}}a_N(d,x) \,=\, a_N(d_{\max},x) \,=\,
a_N\big(f_-(x)\,N+\mathcal{O}(\sqrt{N}\,)\,,\,x\big) \\
&=\, \frac {N!\ \,(2 N)^{-f(x)\,N+\mathcal{O}(\sqrt{N}\,)}\ \,x^{N-2\,f(x)\,N+\mathcal{O}(\sqrt{N}\,)}}
{\big(f(x)\,N+\mathcal{O}(\sqrt{N}\,)\big)!\ \,\big(N-2\,f(x)\,N+\mathcal{O}(\sqrt{N}\,)\big)!}
\end{split}\]
where $f(x):=f_-(x)$. Set also $g(x):=1-2\,f(x)$.
Take the logarithm, divide by $N$ and use the Stirling formula (in the form $\log(n!)=n\,\log n-n+\mathcal{O}(\log n)$ as $n\to\infty$) to find for $N\to\infty$
%\[\begin{split}
%&\frac{\log M_N(x)}{N} \,= \\
%&=\, \frac{N\log N-N}{N}\, - \frac{f(x)N\,\log\big(f(x)N\big)-f(x)N}{N}\, - \frac{g(x)N\,\log\big(g(x)N\big)-g(x)N}{N}\, + \\ &\quad\, - \frac{f(x)N\log(2N)}{N}\, + \frac{g(x)N\log x}{N}\, + \mathcal{O}\big(\frac{\log N}{\sqrt{N}}\,\big)
%\end{split}\]
%simplifying $N$ and isolating first $\log N$ and then $f(x)$ and $g(x)$, one obtains
\[\begin{split}
\frac{\log M_N(x)}{N} =\,&
\big(1-f(x)-g(x)-f(x)\big)\log N\ + \\
&f(x)\,\big(\!-\!\log f(x)+1-\log 2\big) + g(x)\,\big(\!-\!\log g(x)+1+\log x\big) - 1 + \mathcal{O}\big(\frac{\log N}{\sqrt{N}}\big) \;;
\end{split}\]
notice that the coefficient of $\log N$ is zero, hence
\[ \frac{\log M_N(x)}{N} \,\xrightarrow[N\to\infty]{}\,
f(x)\,\big(\!-\!\log f(x)+1-\log 2\big) + g(x)\,\big(\!-\!\log g(x)+1+\log x\big) - 1 \,.\]
As observed before $\log \Zcl_N(x)/N$ must converge to the same limit and the statement is proved.
\endproof

\begin{rk} \label{rk: rewriting of p}
The limit of the pressure and its derivative admit a simple rewriting, which will be useful in the sequel.
%\begin{gather}
%\pcl(x) \,=\, -\frac{1-g(x)}{2}\,-\frac{1}{2}\,\log(1-g(x)) \,=\, -\frac{1-g(x)}{2}\,-\log g(x)+\log x \;, \label{eq: p classic mean field 2} \\[4pt]
%x\,(\pcl)'(x) = g(x) \;.
%\end{gather}
To find it begin observing that the equation $g(x)=y$ can be solved w.r.t. $\!x$ by a direct computation, so that the function $g$ is invertible on $]0,\infty[$ with inverse function
$ g^{-1}(y) = y/\sqrt{1-y}$ for $0<y<1$.
Choosing $y=g(x)$ it follows that
\begin{equation}\label{eq: identity for g(x)}
x = \frac{g(x)}{\sqrt{1-g(x)}}\ ,\quad\text{i.e. }\ \frac{1}{2}\,\log(1-g(x)) = \log g(x)-\log x \;.
\end{equation}
Remembering that $f=(1-g)/2$ and using identity (\ref{eq: identity for g(x)}), the expression (\ref{eq: p classic mean field 1}) becomes
\begin{equation} \label{eq: rewriting of p(x)} \begin{split}
\pcl(x) \,&=\, -\frac{1}{2}\,(1-g(x)) -\frac{1}{2}\,\log(1-g(x)) \\
&=\, -\frac{1}{2}\,(1-g(x)) -\log g(x)+\log x \;.
\end{split}\end{equation}
%\pcl(x) \,&=\, f(x)\,\big(1-\log f(x)-\log 2\big) + g(x)\,\big(1-\log g(x)+\log x\big) - 1 \\
%&=\, \frac{1-g(x)}{2}\,\big(1-\log(1-g(x))\big) + g(x)\,\big(1-\log g(x)+\log x\big) - 1 \\
%&=\, (1-g(x))\,\big(\frac{1}{2}-\frac{1}{2}\,\log(1-g(x))\big) + g(x)\,\big(1-\frac{1}{2}\,\log(1-g(x))\big) - 1 \\
%&=\, -\,\frac{1-g(x)}{2}\, -\frac{1}{2}\,\log(1-g(x))\;.
Now use the first of these expressions to compute the derivative $(\pcl)'(x) = \frac{g'(x)}{2}\,\frac{2-g(x)}{1-g(x)}\,$.
%\[ (\pcl)'(x) =  -\frac{1}{2}\;\frac{\dd}{\dd x}\,\big(1-g(x)+\log(1-g(x))\big) =\, \frac{g'(x)}{2}\;\frac{2-g(x)}{1-g(x)}\]
Write the derivative of $g$ via its inverse function $g'(x) = \frac{1}{(g^{-1})'(g(x))} = \frac{2\,(1-g(x))^{3/2}}{2-g(x)}\,$.
%\[ \big(g^{-1}\big)'(y) = \frac{2-y}{2\,(1-y)^{3/2}}\ \ \Rightarrow\ \ g'(x) = \frac{1}{(g^{-1})'(g(x))} = \frac{2\,(1-g(x))^{3/2}}{2-g(x)} \;;\]
Therefore, substituting and using again (\ref{eq: identity for g(x)}),
\begin{equation} \label{eq: rewriting of p'(x)}
x\,(\pcl)'(x) = x\,\sqrt{1-g(x)} \,=\, g(x) \;. \qedhere
\end{equation}
\end{rk}
\section{Imitative monomer-dimer model} \label{sec: m-d imitative}
The monomer-dimer model on a graph $G$ is characterised by a topological interaction, that is the hard-core constraint which defines the space of states $\DD_G$ (see definition \ref{def: space of states}).
As proved by Heilmann and Lieb \cite{HL, HLprl} this interaction is not sufficient to originate a phase transition: when the thermodynamic limit of the normalized pressure exists, is has to be an analytic function of the parameters.

Now we will consider also another type of interaction, as described in (\ref{main}): we want that the state of a vertex conditions the state of its neighbours, pushing each other to behave in the same way (\textit{imitative interaction} between sites, {attractive interaction} between particles of the same type). %or in the opposite way (\textit{counter-imitative or repulsive interaction}).

We start making the following

\begin{rk} \label{rk: classical model in hamiltonian form}
The probability measure associated to a  monomer-dimer model on the graph $G=(V,E)$ can be rewritten in the Boltzmann form by the following parametrization of the monomer and dimer weights:
\begin{equation}
x_v = \exp(\hm_v)\,,\quad w_e = \exp(\hd_e)
\end{equation}
with $\hm_v,\, \hd_e\in\R$ for all $v\in V,\,e\in E\,$.
Then it is possible to define the \textit{hamiltonian}
\begin{equation}\label{eq: H classic}
-\Hcl_{G}(D) \,:=\, \sum_{v\in V}\,\hm_v\,\1(v\in\MM(D)) \,+\, \sum_{e\in E}\,\hd_e\,\1(e\in D) \quad\forall\,D\in\DD_G \;,
\end{equation}
where $\1(A)$ is $1$ if $A$ is true and $0$ otherwise, and rewrite the partition function (\ref{eq: Z classic}) as
\[ \Zcl_G \,=\, \sum_{D\in\DD_G} \exp(-\Hcl_{G}(D)) \;. \]
%\textcolor{red}{In this section and the following we use the same notations of the previous section $\Zcl_N,\,\pcl,\,g$ to denote those functions composed with $\exp$.}
\end{rk}

\begin{df} \label{def: imitative m-d}
As usual let $\DD_G$ be the set of all possible dimer configurations on the graph $G\,$.
The \textit{imitative monomer-dimer model} on $G$ is obtained by assigning to each vertex $v\in V$ a monomer external field $\hm_v\in\R$ and assigning to each edge $e\in E$ a dimer eternal field $\hd_e\in\R$, a monomer imitation coefficient $\Jm_e\in\R$, a dimer imitation coefficient $\Jd_e\in\R$ and a counter-imitation coefficient $\Jmd_e\in\R$ and then considering the following probability measure on the set $\DD_G$:
\[ \muim_{G}(D) \,:=\, \frac{1}{\Zim_G}\;\exp(-\Him_{G}(D)) \quad\forall\,D\in\DD_G \;, \]
where the \textit{hamiltonian} is: $\forall\,D\in\DD_G$
\begin{equation}\label{eq: H imitative}\begin{split}
-\Him_{G}(D) \,:=&\,
\sum_{v\in V}\,\hm_v\,\1(v\!\in\!\MM(D)) \,+\, \sum_{uv\in E}\hd_{uv}\,\1(uv\!\in\!D) \ +\\
&\sum_{uv\in E}\Jm_{uv}\,\1(u\!\in\!\MM(D),\,v\!\in\!\MM(D)) \,+\, \sum_{uv\in E}\Jd_{uv}\,\1(u\!\notin\!\MM(D),\,v\!\notin\!\MM(D)) \ +\\
&\sum_{uv\in E}\Jmd_{uv}\,[\1(u\!\in\!\MM(D),\,v\!\notin\!\MM(D))+\1(u\!\notin\!\MM(D),\,v\!\in\!\MM(D))]
\end{split}\end{equation}
and the \textit{partition function} is $\Zim_G:= \sum_{D\in\DD_G}\exp(-\Him_{G}(D))\,$.
As usual $\log\Zim_G$ is called \textit{pressure}.
\end{df}

\begin{rk} \label{rk: imitative monomer density}
With uniform monomer field $\hm_v\equiv\hm$, the \textit{monomer density}, i.e. the expected fraction of monomers on the graph, in the imitative model is the derivative of the pressure w.r.t. $\hm\,$:
\[ \mim_G:=\sum_{D\in\DD_G}\frac{|\MM(D)|}{|V|}\; \muim_{G}(D) \,=\, \frac{\partial}{\partial\hm}\,\frac{\log \Zim_G}{|V|} \;.\]
\end{rk}

In the following remark we show the imitative monomer-dimer model, under the hypothesis of uniform dimer field, depends only on 2 families of parameters (while a priori we introduced 5 families).
Moreover we show that the imitative monomer-dimer model is related to the Ising model, but it is not trivially equivalent to it because of the topological lack of symmetry between monomers and dimers.

\begin{rk} \label{rk: H imitative alpha sigma}
Set $\alpha_v(D):=\1(v\in\MM(D))$.
Notice that in the hamiltonian (\ref{eq: H imitative}) the only functions of the dimer configuration $D$ that can not be expressed in terms of the $\{\alpha_v\}_{v\in V}$ are the $\{\1(uv\in D)\}_{uv\in E}$; indeed, given the configuration of monomers, the configuration of dimers in general is not determined in a unique way.\\
But if we consider only uniform dimer field $\hd_{uv}\equiv\hd$, using the identities $|D|=\frac{|V|-|\MM(D)|}{2}=\frac{1}{2}(|V|-\sum_v\alpha_v(D))$, $\1(u\!\in\!\MM,v\!\in\!\MM)=\alpha_u\alpha_v$, $\1(u\!\notin\!\MM,v\!\notin\!\MM)=(1-\alpha_u)(1-\alpha_v)=1-\alpha_u-\alpha_v+\alpha_u\alpha_v$, $\1(u\!\in\!\MM,v\!\notin\!\MM)=\alpha_u(1-\alpha_v)=\alpha_u-\alpha_u\alpha_v$ we obtain:
\begin{equation}\label{eq: H imitative alpha}
-\Him_G(D) \,=\,
%\sum_{v\in V}\hm_v\alpha_v(D) \,+\, \hd|D| \,+ \sum_{uv\in E}\!\Jm_{uv}\alpha_u(D)\alpha_v(D) \,+ \sum_{uv\in E}\!\Jd_{uv}(1-\alpha_u(D))(1-\alpha_v(D)) \\
C' \,+ \sum_{v\in V}h'_v\,\alpha_v(D) \,+ \sum_{uv\in E}J'_{uv}\,\alpha_u(D)\,\alpha_v(D)\;,
\end{equation}
where we set:
\[\begin{split}
&h'_v:=\hm_v-\frac{1}{2}\hd-\sum_{u\sim v}\Jd_{uv}+\sum_{u\sim v}\Jmd_{uv}\,,\quad
J'_{uv}:=\Jm_{uv}+\Jd_{uv}-2\Jmd_{uv}\,,\\
&C':=\frac{1}{2}\hd|V|+\sum_{uv\in E}\Jd_{uv}\;.
\end{split}\]
%Notice that if original coefficients are uniform $\hd_e\equiv\hd$, $\hm_v\equiv\hm$, $\Jm_e\equiv\Jm$, $\Jd_e\equiv\Jd$ and the graph is regular of degree $r$, then also the new coefficients are uniform, precisely
%\[ h_v\equiv h=\hm-\frac{1}{2}\hd+r\Jd\,,\quad J_{uv}\equiv J=\Jm+\Jd \;.\]
Now set $\sigma_v(D):=2\,\alpha_v(D)-1\in\{-1,1\}$.
To draw a parallel with the Ising model, we can rewrite the hamiltonian (\ref{eq: H imitative alpha}) as a function of $\{\sigma_v\}_{v\in V}$. Using $\alpha_v=\frac{1}{2}(\sigma_v+1)$, $\alpha_u\alpha_v=\frac{1}{4}(\sigma_u\sigma_v+\sigma_u+\sigma_v+1)$, we obtain:
\begin{equation}\label{eq: H imitative sigma}
-\Him_G(D) \,=\,
%\sum_{v\in V}\hm_v\alpha_v(D) \,+\, \hd|D| \,+ \sum_{uv\in E}\!\Jm_{uv}\alpha_u(D)\alpha_v(D) \,+ \sum_{uv\in E}\!\Jd_{uv}(1-\alpha_u(D))(1-\alpha_v(D)) \\
C'' \,+ \sum_{v\in V}h''_v\,\sigma_v(D) \,+ \sum_{uv\in E}J''_{uv}\,\sigma_u(D)\,\sigma_v(D)\;,
\end{equation}
where we set:
\begin{gather*}
h''_v \,:=\, \frac{1}{2}h'_v+\frac{1}{4}\sum_{u\sim v}J'_{uv} \,=\,
\frac{1}{2}\hm_v-\frac{1}{4}\hd+\frac{1}{4}\sum_{u\sim v}\Jm_{uv}-\frac{1}{4}\sum_{u\sim v}\Jd_{uv}  \;,\\
J''_{uv} \,:=\, \frac{1}{4}J'_{uv} \,=\, \frac{1}{4}\Jm_{uv}+\frac{1}{4}\Jd_{uv}-\frac{1}{2}\Jmd_{uv} \;,\\
C'' \,:=\, C+\frac{1}{2}\sum_{v\in V}h'_v+\frac{1}{4}\sum_{uv\in E}J'_{uv} \,=\, \frac{1}{2}\sum_{v\in V}\hm_v + \frac{1}{4}\hd|V| + \frac{1}{4}\sum_{uv\in E}\Jm_{uv} + \frac{1}{4}\sum_{uv\in E}\Jd_{uv} \;.
\end{gather*}
%where we set:
%\[ h''_v:=\frac{1}{2}h'_v+\frac{1}{4}\sum_{u\sim v}J'_{uv},\quad
%J''_{uv}:=\frac{1}{4}J'_{uv}\,,\quad
%C'':=C+\frac{1}{2}\sum_{v\in V}h'_v+\frac{1}{4}\sum_{uv\in E}J'_{uv}\;. \]
%Substituting the expression of $h'_v,\,J'_{uv},\,C'$, we obtain:
%\[\begin{split}
%&h''_v:=\frac{1}{2}\hm_v-\frac{1}{4}\hd+\frac{1}{4}\sum_{u\sim v}\Jm_{uv}-\frac{1}{4}\sum_{u\sim v}\Jd_{uv}\,,\quad
%J''_{uv}:=\frac{1}{4}\Jm_{uv}+\frac{1}{4}\Jd_{uv}-\frac{1}{2}\Jmd_{uv}\,,\quad\\
%&C'':=\frac{1}{2}\sum_{v\in V}\hm_v + \frac{1}{4}\hd|V| + \frac{1}{4}\sum_{uv\in E}\Jm_{uv} + \frac{1}{4}\sum_{uv\in E}\Jd_{uv} \;.
%\end{split}\]
%Notice that if original coefficients are uniform $\hd_e\equiv\hd$, $\hm_v\equiv\hm$, $\Jm_e\equiv\Jm$, $\Jd_e\equiv\Jd$ and the graph is regular of degree $r$, then also these new coefficients are uniform, precisely
%\[ h'_v\equiv h'=\frac{1}{2}\hm-\frac{1}{4}\hd+\frac{r}{4}\Jm+\frac{3r}{4}\Jd\,,\quad J'_{uv}\equiv J'=\frac{1}{4}\Jm+\frac{1}{4}\Jd \;.\]
Now consider the usual hamiltonian of the Ising model on the graph $G$
\[ -\Hising_G(\sigma) \,:= \sum_{v\in V}h''_v\,\sigma_v \,+ \sum_{uv\in E}J''_{uv}\,\sigma_u\,\sigma_v \quad\forall\,\sigma\in\{-1,1\}^V \;.\]
%the associated partition function $\Zising_G = \sum_{\sigma\in\{\pm1\}^V}\exp(-\Hising_G(\sigma))$, probability measure $\mu^{\textup{\tiny{ISING}}}_G(\sigma) = \exp(-\Hising_G(\sigma))/\Zising_G$ and expected value $\langle f\rangle^{\textup{\tiny{ISING}}}_G = \sum_{\sigma\in\{\pm1\}^V}f(\sigma)\,\mu^{\textup{\tiny{ISING}}}_G(\sigma)$.
From identity (\ref{eq: H imitative sigma}), it follows immediately that
\[ \Zim_G \,= \sum_{D\in\DD_G}\exp(-\Him_G(D)) \,= \sum_{\sigma\in\{\pm1\}^V}\!\!\Card\{D\in\DD_G,\,\sigma(D)=\sigma\}\,\exp(-\Hising_G(\sigma))\, e^{C''} \;, \]
that is, setting $\nu(\sigma):=\Card\{D\in\DD_G,\,\sigma(D)=\sigma\}=$ number of possible dimer configurations with positions of the monomers given by the $1$'s in $\sigma$,
\begin{equation}
\Zim_G \,=\, e^{C''}\,\Zising_G\;\langle\,\nu\,\rangle_{G}^{\textup{\tiny{ISING}}} \;,
\end{equation}
where $\Zising_G := \sum_{\sigma\in\{\pm1\}^V}e^{-\Hising_G(\sigma)}$ and $\langle f\rangle^{\textup{\tiny{ISING}}}_G := \sum_{\sigma\in\{\pm1\}^V}f(\sigma)\,e^{-\Hising_G(\sigma)}/\Zising_G\,$.\\[2pt]
We will see that in the case of complete graph the correct normalisation gives to the parameters $C''$ and $h''$ a non trivial dependence on the volume, which can be viewed as the effect of the hard core interaction on the entropy of the system and shows that the exact solution we are about to derive cannot be trivially related to the mean-field ferromagnet.
\end{rk}

\subsection{Imitative monomer-dimer model on the complete graph}

Now we study the imitative model on the complete graph $K_N=(V_N,E_N)$ with uniform parameters $\hm_v\equiv\hm$, $\hd_e\equiv\hd$, $\Jm_e\equiv\Jm$, $\Jd_e\equiv\Jd$, $\Jmd_e\equiv\Jmd$ for all $v\in V_N,\,e\in E_N$.\\
Remember that the correct normalisation for the monomer dimer model is given by the dimer weight $w/N$, that is dimer field $\hd-\log N$. Further for the imitative model we will see that the normalisations $\Jm/N,\,\Jd/N,\,\Jmd/N$ are also required.
Hence we consider the following hamiltonian: $\forall\,D\in\DD_{K_N}$
\begin{equation} \label{eq: H imitative complete graph} \begin{split}
-\Him_N(D) :=\,
&\hm\sum_{v\in V}\1(v\!\in\!\MM(D)) \,+\, (\hd-\log N)\sum_{uv\in E}\1(uv\!\in\!D) \ +\\
&\frac{\Jm}{N}\sum_{uv\in E}\1(u\!\in\!\MM(D),\,v\!\in\!\MM(D)) \,+\, \frac{\Jd}{N}\sum_{uv\in E}\1(u\!\notin\!\MM(D),\,v\!\notin\!\MM(D)) \ +\\
&\frac{\Jmd}{N}\sum_{uv\in E}[\1(u\!\in\!\MM(D),\,v\!\notin\!\MM(D))+\1(u\!\notin\!\MM(D),\,v\!\in\!\MM(D))]
\end{split}\end{equation}
and the associated partition function $\Zim_N \,:=\, \sum_{D\in\DD_G}\exp(-\Him_{N}(D))\,$.
%In $\Him_N(D)$ the dependence on $\hm,\hd,\Jm,\Jd$ is kept implicit to shorten the notation.

\begin{rk} \label{rk: H imitative complete graph order parameter}
Given a dimer configuration $D$ on the graph $K_N$, denote the fraction of vertices covered by monomers by
\[ m_N(D):=\frac{|\MM(D)|}{N}\ \in[0,1] \;.\]
On the complete graph the hamiltonian (\ref{eq: H imitative complete graph}) of the imitative model admits a useful rewriting, which shows that it depends on a dimer configuration $D$ only via the quantity $m_N(D)$. Precisely: $\forall D\in\DD_{K_N}$
\begin{equation} \label{eq: imitative H complete graph rewriting} \begin{split}
-\frac{1}{N}\,\Him_{N}(D) \,=\, a\;m_N(D)^2 \,+\, b_N\;m_N(D) \,+\, c_N
\end{split}\end{equation}
with
\begin{gather*}
a \,:=\, \frac{1}{2}(\Jm+\Jd-2\Jmd)\;,\\[1pt]
b_N \,:=\, \frac{\log N}{2} + \hm-\frac{\hd}{2} - \frac{N-1}{2N}(\Jd-\Jmd) - \frac{1}{2N}(\Jm+\Jd-2\Jmd) \;,\\[3pt]
c_N \,:=\, -\frac{\log N}{2} + \frac{\hd}{2} + \frac{N-1}{2N}\Jd \;.
\end{gather*}
To prove it, it suffices to rewrite the hamiltonian (\ref{eq: H imitative complete graph}) as in expression (\ref{eq: H imitative alpha}) and then observe that on the complete graph $\frac{1}{N}\sum_{v\in V_N}\alpha_v=m_N$, $\frac{1}{N}\sum_{uv\in E_N}\alpha_u\alpha_v=\frac{1}{2}\,N\,m_N^2-\frac{1}{2}\,m_N$.
\end{rk}

\begin{rk} \label{rk: mean field classical limit h}
We need to re-state the results of Section \ref{sec: m-d classic} using the hamiltonian form introduced in this section.
The partition function $\Zcl_N(x)$ of the  monomer-dimer model on the complete graph defined by (\ref{eq: Z classic complete graph}) can be rewritten with a slight abuse of notation as
\[\begin{split}
\Zcl_N(h) &= \sum_{D\in\DD_{K_N}} \exp\big(\,h\,|\MM(D)| -\log N\;|D|\,\big) \\
&= \sum_{D\in\DD_{K_N}} \exp N \big(\,(h+\frac{1}{2}\log N)\,m_N(D) - \frac{1}{2}\log N\,\big)\;,
\end{split}\]
where the monomer and dimer weights have been rewritten as $x=e^h$, $w=1/N=e^{-\log N}$.
Using this notation proposition \ref{prop: mean field classic limit} and remark \ref{rk: rewriting of p} can be re-stated as follows.
The pressure per particle on the complete graph admits thermodynamic limit:
\[ \forall\,h\in\R \quad\exists\;\lim_{N\to\infty}\frac{\log \Zcl_N(h)}{N} \,=\, \pcl(h) \]
where $\pcl$ is an analytic function of $h$, precisely:
\begin{gather}
\pcl(h) \,:=\, -\frac{1-g(h)}{2}-\frac{1}{2}\log(1-g(h)) \,=\, -\frac{1-g(h)}{2}-\log g(h)+ h \label{eq: p(h)}\\[2pt]
g(h) \,:=\, \frac{1}{2}\,(\sqrt{e^{4h}+4\,e^{2h}}-e^{2h}) \;. \label{eq: g(h)}
\end{gather}
Note that, since $h\mapsto\frac{\log \Zcl_N(h)}{N}$ is a convex function and its limit $\pcl$ is differentiable, also the monomer density (see remark \ref{rk: monomer density}) converges, and precisely
\[ \mcl_N \,=\, \frac{\partial}{\partial h} \frac{\log \Zcl_N}{N}\ \xrightarrow[N\to\infty]{}\ (\pcl)' \,=\, g \,.\]
The properties of this function $g$ which will be needed in Section \ref{sec: analysis} are studied in the Appendix.
\end{rk}

Thank to the previous remarks, in the case $\Jm+\Jd-2\Jmd>0$ the imitative model can be exactly solved.
Our technique is the same used by Guerra\cite{G} to solve the ferromagnetic Ising model on the complete graph.

\begin{thm} \label{thm: imitative mean field limit}
Let $\hm,\,\hd,\,\Jm,\,\Jd,\,\Jmd\in\R$ such that $\Jm+\Jd-2\Jmd\geq0$.
The pressure per particle of the imitative monomer-dimer model on the complete graph defined by hamiltonian (\ref{eq: H imitative complete graph}) admits thermodynamic limit:
\[ \exists\;\lim_{N\to\infty}\frac{\log\Zim_N}{N} \,=:\, \pim \;\in\R \;.\]
This limit satisfies a variational principle:
\[ \pim \,=\, \sup_{m}\,\vpim(m) \;,\]
where the $\sup$ can be taken indifferently over $m\in[0,1]$ or $m\in\R$, and
\[\begin{split}
\vpim(m) \,:=\, &-\frac{1}{2}(\Jm+\Jd-2\Jmd)\;m^2 \,+\, \frac{1}{2}(\hd+\Jd) \ +\\
& +\, \pcl\big(\,(\Jm+\Jd-2\Jmd)\,m\,+\hm-\frac{1}{2}\hd-\Jd+\Jmd\,\big)
\end{split}\]
where the function $\pcl$ is defined by (\ref{eq: p(h)}), (\ref{eq: g(h)}).
\end{thm}

\proof
The proof is done providing a lower and an upper bound for the pressure per particle.\\[2pt]
$\mathbf{[LowerBound]}$ Fix $m\in\R$. As $\big(m_N(D)-m)^2\geq0$, clearly $m_N(D)^2\geq2\,m\,m_N(D)-m^2$.
Hence by remark \ref{rk: H imitative complete graph order parameter}, using that by hypothesis $a\geq0$,
\[\begin{split}
-\Him_{N}(D) \,&=\, N\,\big(a\,m_N(D)^2\,+\,b_N\,m_N(D)\,+\,c_N\big) \,\geq\\
&\geq\, N\,\big((2\,a\,m+b_N)\,m_N(D)\,-a\,m^2+c_N\big)
\end{split}\]
thus
\[\begin{split}
\Zim_N \,&=\, \sum_{D}\exp(-\Him_{N}(D)) \,\geq\,
\sum_{D}\exp N\big((2\,a\,m+b_N)\,m_N(D)-a\,m^2+c_N\big) \,=\\
&=\, e^{N\,\gamma_N(m)}\ \Zcl_N\big(\alpha_N(m)\big)
\end{split}\]
where the last equality is due to remark \ref{rk: mean field classical limit h} and $\gamma_N(m) := -\frac{1}{2}(\Jm+\Jd-2\Jmd)\,m^2 + \frac{1}{2}\hd + \frac{N-1}{2N}\Jd$ and $\alpha_N(m) := (\Jm+\Jd-2\Jmd)\,m+\hm-\frac{\hd}{2}-\frac{N-1}{N}\,(\Jd-\Jmd)-\frac{1}{2N}(\Jm+\Jd-2\Jmd)\,$.\\[2pt]
$\mathbf{[UpperBound]}$ Set $\mathcal{A}_N:=\textrm{Im}(m_N) = \{0,\frac{1}{N},\dots,\frac{N-1}{N},1\}$. Clearly, writing $\delta$ for the Kronecker delta, $\sum_{m\in\mathcal{A}_N}\delta_{m,m_N(D)} = 1$ and $F(m_N(D)^2)\,\delta_{m,m_N(D)} = F(2\,m\,m_N(D)-m^2)\,\delta_{m,m_N(D)}$ for any real function $F$.
Hence by remark \ref{rk: H imitative complete graph order parameter},
\[\begin{split}
\delta_{m,m_N(D)}\,\exp(-\Him_{N}(D)) \,&=\, \delta_{m,m_N(D)}\,\exp N(a\,m_N(D)^2+b_N\,m_N(D)+c_N) \,=\\
&=\, \delta_{m,m_N(D)}\,\exp N\big((2\,a\,m+b_N)\,m_N(D)\,-a\,m^2+c_N\big)
\end{split}\]
thus
\[\begin{split}
\Zim_N \,&=\,
\sum_{D} \sum_{m\in\mathcal{A}_N} \delta_{m,m_N(D)}\;\exp(-\Him_{N}(D)) \,=\\
&=\, \sum_{D} \sum_{m\in\mathcal{A}_N} \delta_{m,m_N(D)}\; \exp N\big((2\,a\,m+b_N)\,m_N(D)\,-a\,m^2+c_N\big) \,\leq\\
&\leq \sum_{m\in\mathcal{A}_N} \sum_{D}\; \exp N\big((2\,a\,m+b_N)\,m_N(D)\,-a\,m^2+c_N\big)\,=\\
&= \sum_{m\in\mathcal{A}_N} e^{N\,\gamma_N(m)}\ \Zcl_N\big(\alpha_N(m)\big) \,\leq\,
(N+1)\sup_{m\in[0,1]}\big\{e^{N\,\gamma_N(m)}\ \Zcl_N\big(\alpha_N(m)\big)\big\} \;.
\end{split}\]
Therefore putting together lower and upper bound we have found:
\[ \sup_{m\in[0,1]}\big\{e^{N\,\gamma_N(m)}\ \Zcl_N\big(\alpha_N(m)\big)\big\} \,\leq\, \Zim_N \,\leq\, (N+1)\sup_{m\in[0,1]}\big\{e^{N\,\gamma_N(m)}\ \Zcl_N\big(\alpha_N(m)\big)\big\} \;.\]
Then, taking the logarithm and dividing by $N$,
\[ 0 \,\leq\, \frac{\log\Zim_N}{N} \;- \sup_{m\in[0,1]}\big\{\gamma_N(m) \,+\, \frac{\log \Zcl_N\big(\alpha_N(m)\big)}{N}\big\} \,\leq\, \frac{\log(N+1)}{N} \,\xrightarrow[N\to\infty]{}\,0 \;.\]
Now for any $N\in\N$ the pressure $h\mapsto\frac{\log \Zcl_N(h)}{N}$ is a convex function, hence
% una successione puntualmente convergente di funzioni convesse è anche uniformemente convergente sui compatti
as $N\to\infty$ the convergence $\frac{\log \Zcl_N(h)}{N} \rightarrow \pcl(h)$ of remark \ref{rk: mean field classical limit h} is uniform in $h$ on compact sets.\\
Moreover notice that as $N\to\infty$, $\alpha_N(m) \rightarrow \alpha(m) := (\Jm+\Jd-2\Jmd)\,m+\hm-\frac{\hd}{2}-\Jd+\Jmd$ and $\gamma_N(m) \rightarrow \gamma(m) := -\frac{1}{2}(\Jm+\Jd-2\Jmd)\,m^2+\frac{1}{2}\hd+\frac{1}{2}\Jd$ uniformly in $m$.
Therefore, exploiting also the fact that $\pcl$ is lipschitz,
% se gn ed fn convergono uniformemente a g ed f risp., se inoltre g è lipschitziana, allora anche la composizione gn(fn) converge uniformemente a g(f)
\[ \gamma_N(m) \,+\, \frac{\log \Zcl_N\big(\alpha_N(m)\big)}{N}\ \xrightarrow[N\to\infty]{}\
\gamma(m) \,+\, \pcl\big(\alpha(m)\big) \]
where the convergence is uniform in $m$ on compact sets.
As a consequence also
% se fn converge uniformemente ad f, allora sup(fn) converge a sup(f)
\[ \sup_{m\in[0,1]}\big\{\gamma_N(m) \,+\, \frac{\log \Zcl_N\big(\alpha_N(m)\big)}{N}\big\}\ \xrightarrow[N\to\infty]{}\
\sup_{m\in[0,1]}\big\{\gamma(m) \,+\, \pcl\big(\alpha(m)\big)\big\} \;.\]
This concludes the proof.
\endproof
\section{Analysis of the solution of the imitative monomer-dimer model on the complete graph} \label{sec: analysis}
In this section we study the properties of the solution given by theorem \ref{thm: imitative mean field limit}.
We set $\hm=:h\,$, $\hd=0\,$, $\Jm=\Jd=:J>0\,$, $\Jmd=0$ in (\ref{eq: H imitative complete graph}); that is we consider the hamiltonian
\begin{equation} \label{eq: imitative H complete graph 2param} \begin{split}
-\Him_N(D) \,:=\ &
h \sum_{v\in V_N}\1(v\!\in\!\MM(D)) \,-\, \log N \!\sum_{uv\in E_N}\!\1(uv\!\in\!D) \ +\\
&\frac{J}{N} \sum_{uv\in E_N}\!\big[\,\1(u\!\in\!\MM(D),v\!\in\!\MM(D)) \,+\, \1(u\!\notin\!\MM(D),v\!\notin\!\MM(D))\,\big] \;.
\end{split}\end{equation}
This choice can be done without loss of generality. Indeed, as shown by remark \ref{rk: H imitative alpha sigma}, the general hamiltonian (\ref{eq: H imitative complete graph}) rewrites as $h'\sum_{v\in V_N}\1(v\!\in\!\MM)-\log N\sum_{uv\in E_N}\1(uv\!\in\!D)+J'/N\sum_{uv\in E_N}\1(u\!\in\!\MM,v\!\in\!\MM)$, up to a constant, for suitable $h',J'$. Now applying the invertible linear change of parameters $J'=2\,J$, $h'=h-J$, we obtain the hamiltonian (\ref{eq: imitative H complete graph 2param}).\\[2pt]
The associated partition function is denoted $\Zim_N(h,J)$. By theorem \ref{thm: imitative mean field limit}
\[ \frac{\log\Zim_N(h,J)}{N}\ \xrightarrow[N\to\infty]{}\ \pim(h,J) \,=\, \sup_{m}\,\vpim(m,h,J) \]
where the $\sup$ can be taken indifferently over $m\in[0,1]$ or $m\in\R$, and
\begin{equation} \label{eq: vp(m,h,J)}
\vpim(m,h,J) \,:=\, -J\,m^2 \,+\, \frac{J}{2} \,+\, \pcl\big((2m-1)J+h\big)
\end{equation}
with the analytic function $\pcl$ defined by (\ref{eq: p(h)}), (\ref{eq: g(h)}).\\
Thus we want to study the following variational problem:
\[ \textit{maximize } \vpim(m,h,J) \textit{ with respect to } m\in[0,1]\, (\in\R) \]
and in particular we are interested in the value(s) of $m=m^*(h,J)\in[0,1]$ where the maximum is reached, because of its physical meaning that we will explain in remark \ref{rk: m*=monomer density}.

\begin{rk} \label{rk: extremum iff ce}
Remembering that $(\pcl)'=g$, one computes
\begin{gather}
\frac{\partial\vpim}{\partial m}\,(m,h,J) \,=\, -2J\,m \,+\, 2J\;g\big((2m-1)J+h\big) \label{eq: deriv vp w.r.t. m}\\
\frac{\partial^2\vpim}{\partial m^2}\,(m,h,J) \,=\, -2J \,+\, (2J)^2\;g'\big((2m-1)J+h\big) \label{eq: second deriv vp w.r.t. m}
\end{gather}
Since $0<g<1$, it follows that for every $J>0,\,h\in\R$
\begin{equation} \label{eq: vp incr m<=0, decr m>=1}
\frac{\partial\vpim}{\partial m}\,(m,h,J)>0 \quad\forall\,m\in\,]-\infty,0] \;, \quad
\frac{\partial\vpim}{\partial m}\,(m,h,J)<0 \quad\forall\,m\in[1,\infty[ \;.
\end{equation}
Therefore $\vpim(\cdot\,,h,J)$ attains its maximum in (at least) one point $m=m^*(h,J)\in\,]0,1[\,$, which satisfies
\begin{gather}
\frac{\partial\vpim}{\partial m}\,(m,h,J) = 0 \qquad\text{i.e.}\quad m = g\big((2m-1)J+h\big) \;, \label{eq: ce} \\
\frac{\partial^2\vpim}{\partial m^2}\,(m,h,J ) \leq 0 \qquad\text{i.e.}\quad g'\big((2m-1)J+h\big) \leq\frac{1}{2J} \label{eq: ineq}\;.
\end{gather}
\end{rk}

The following remark explains the physical meaning of the maximum point $m^*$.

\begin{rk} \label{rk: m*=monomer density}
Let $m^*(h,J)$ denote a point maximizing the function $m\mapsto\vpim(m,h,J)$ on $[0,1]$, that is
\[ \pim(h,J) \,=\, \vpim(m^*(h,J),h,J) \,.\]
\textit{Assume} the function $h \mapsto m^*(h,J)$ is differentiable.
Then $h \mapsto \pim(h,J)$ is differentiable and, using equation (\ref{eq: ce}) for $m^*(h,J)$, identity (\ref{eq: vp(m,h,J)}) and $(\pcl)'=g$, one finds
\begin{equation} \label{eq: deriv pi w.r.t. h}
\frac{\partial \pim}{\partial h}(h,J) \,=\, m^*(h,J) \;.
\end{equation}
%\begin{equation} \label{eq: deriv pi w.r.t. h}
%\frac{\partial \pim}{\partial h} \,=\,
%\underbrace{\frac{\partial \vpim}{\partial m}(m^*,h,J)}_{=\,0} \, \frac{\partial m^*}{\partial h} + \frac{\partial \vpim}{\partial h} \,=\, (\pcl)'((2m^*-1)J+h) \,=\, g((2m^*-1)J+h) \,=\, m^*
%\end{equation}
In other terms $m^*$ is the thermodynamic limit of the monomer density of the imitative monomer-dimer model on the complete graph (see remark \ref{rk: imitative monomer density}).
Indeed by theorem \ref{thm: imitative mean field limit}, exploiting convexity of the function $h\mapsto\frac{\log\Zim_N}{N}\,$,
\[ \mim_N \,=\, \frac{\partial}{\partial h}\frac{\log\Zim_N}{N}\ \xrightarrow[N\to\infty]{}\ \frac{\partial \pim}{\partial h} \,=\, m^* \;.\]
\end{rk}
\subsection{Solutions of the consistency equation $m=g\big((2m-1)J+h\big)$: classification, regularity properties, asymptotic behaviour.}
As a first step we study all the \textit{stationary points} of the function $m\mapsto\vpim(m,h,J)$: by remark \ref{rk: extremum iff ce} one of them will be the \textit{global maximum point} we are interested in.

The stationary points are characterized by equation (\ref{eq: ce}), which can not be explicitly solved.
Anyway their number and a rough approximation of their values can be determined by studying inequality (\ref{eq: ineq}), which admits explicit solution.

The next proposition displays the intervals of concavity/convexity of the function $m\mapsto\vpim(m,h,J)$. Set
\begin{equation} \label{eq: Jc}
J_c := \frac{1}{4\,(3-2\sqrt{2})} \,\approx\, 1.4571 \;.
\end{equation}

\begin{prop} \label{prop: concavity/convexity of vp}
For $0<J<J_c$ and $h\in\R$
\[ \frac{\partial^2 \vpim}{\partial m^2}\,(m,h,J) \,<\, 0 \quad\forall\,m\in\R \;.\]
For $J\geq J_c$ and $h\in\R$
\[ \frac{\partial^2 \vpim}{\partial m^2}\,(m,h,J)
\,\begin{cases}
\,< 0 & \text{iff }\,m<\phi_1(h,J)\text{ or }\,m>\phi_2(h,J) \\[2pt]
\,> 0 & \text{iff }\,\phi_1(h,J)<m<\phi_2(h,J)
\end{cases} \;,\]
where for $i=1,2$
%\begin{gather}
%\phi_i(h,J) \,:=\, \frac{1}{2} - \frac{h}{2J} + \frac{1}{4J}\,\log a_i(J) \;, \label{eq: phi(h,J)}\\
%a_1(J) \,:=\, 2J - 2 - \frac{1}{8J} - (J - \frac{1}{4})\, \sqrt{4 - \frac{6}{J} + \frac{1}{4J^2}} \\
%a_2(J) \,:=\, 2J - 2 - \frac{1}{8J} + (J - \frac{1}{4})\, \sqrt{4 - \frac{6}{J} + \frac{1}{4J^2}} \ .
%\end{gather}
\begin{gather}
\phi_i(h,J) \,:=\, \frac{1}{2} - \frac{h}{2J} + \frac{1}{4J}\,\log a_i(J) \;, \label{eq: phi(h,J)} \\[4pt]
a_{1,2}(J) \,:=\, \frac{-(\frac{1}{(2J)^2}+\frac{8}{2J}-4)\, \mp \,(2-\frac{1}{2J})\sqrt{\frac{1}{(2J)^2}-\frac{12}{2J}+4}}{\frac{4}{2J}} \label{eq: a1,a2(J)}
\end{gather}
%$$\alpha_{\pm}(c) := \frac{-(c^2+8c-4)\,\pm\,(2-c)\sqrt{c^2-12c+4}}{4\,c}$$
Observe that $\phi_1(h,J)\leq\phi_2(h,J)$ for all $h\in\R,\,J\geq J_c$ and equality holds iff $J=J_c$ (since $a_1(J_c)=a_2(J_c)$).
\end{prop}

\proof
It follows from the expression (\ref{eq: second deriv vp w.r.t. m}) through a direct computation done in lemma \ref{lemA: g'<c} of the Appendix, taking $\xi=(2m-1)J+h$ and $c=\frac{1}{2J}\,$.
\endproof

Using the previous proposition we can determine \textit{how many}, of \textit{what kind} and \textit{where} the stationary points of $\vpim(\cdot\,,h,J)$ are.

\begin{prop}[Classification] \label{prop: stationary points of vp}
The equation (\ref{eq: ce}) in $m$ has the following properties:
\begin{itemize}
\item[1.] If $0<J\leq J_c$ and $h\in\R$, there exists only one solution $m(h,J)$.
It is the maximum point of $\vpim(\cdot\,,h,J)$.
\item[2.] If $J>J_c$ and $\psi_2(J)<h<\psi_1(J)$, then there exist three solutions $m_1(h,J)$, $m_0(h,J)$, $m_2(h,J)$.
Moreover $m_1(h,J)<\phi_1(h,J)$ and $m_2(h,J)>\phi_2(h,J)$ are two local maximum points, while
$\phi_1(h,J)<m_0(h,J)<\phi_2(h,J)$ is a local minimum point of $\vpim(\cdot\,,h,J)$.
\item[3.] If $J>J_c$ and $h>\psi_1(J)$, there exists only one solution $m_2(h,J)$.
Moreover $m_2(h,J)>\phi_2(h,J)$ and it is the maximum point of $\vpim(\cdot\,,h,J)$.
\item[4.] If $J>J_c$ and $h=\psi_1(J)$, there exist two solution $m_1(h,J)$, $m_2(h,J)\,$.
Moreover $m_1(h,J)=\phi_1(h,J)$ is a point of inflection, while
$m_2(h,J)>\phi_2(h,J)$ is the maximum point of $\vpim(\cdot\,,h,J)$.
\item[5.] If $J>J_c$ and $h<\psi_2(J)$, there exists only one solution $m_1(h,J)$.
Moreover $m_1(h,J)<\phi_1(h,J)$ and it is the maximum point of $\vpim(\cdot\,,h,J)$.
\item[6.] If $J>J_c$ and $h=\psi_2(J)$, there exist two solutions $m_1(h,J)$, $m_2(h,J)\,$.
Moreover $m_2(h,J)=\phi_2(h,J)$ is a point of inflection, while
$m_1(h,J)<\phi_1(h,J)$ is the maximum point of $\vpim(\cdot\,,h,J)$.
\end{itemize}
\vspace{6pt}
Here $\phi_1,\,\phi_2$ are defined by (\ref{eq: phi(h,J)}), while for $i=1,2$ and $J\geq J_c$
%\begin{equation}
%\psi_i(J) \,:=\, J + \frac{1}{2}\,\log a_i(J) - J\,\big(\sqrt{a_i(J)^2+4\,a_i(J)}\,-a_i(J)\,\big) \;.
%\end{equation}
\begin{equation} \label{eq: psi(J)}
\psi_i(J) \,:=\, J + \frac{1}{2}\log a_i(J) - 2J\, g\big(\frac{1}{2}\log a_i(J)\big) \;,
\end{equation}
where $a_i$ and $g$ are defined respectively by (\ref{eq: a1,a2(J)}) and (\ref{eq: g(h)}).
Observe that $\psi_2(J)\leq\psi_1(J)$ for all $J\geq J_c$ and equality holds iff $J=J_c$.
\end{prop}

\begin{figure}[h]
\centering
\includegraphics{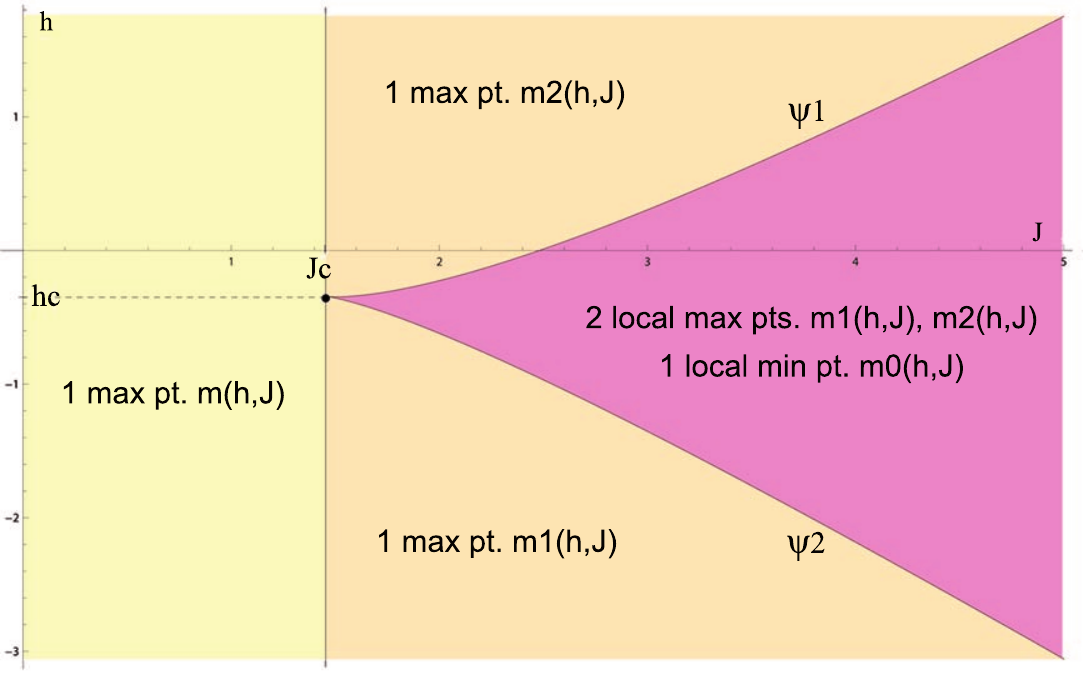}
\caption{Number and nature of the stationary points of the function $m\mapsto\vpim(m,h,J)$ in the regions of the plane $(h,J)$.}
\label{fig: stationary points regions}
\end{figure}

\proof
Fix $h\in\R,\,J>0$ and to shorten the notation set $G(m):=\frac{\partial\vpim}{\partial m}(m,h,J)$, observing it is a continuous (smooth) function.\\[2pt]
$\bullet$ Suppose $J\leq J_c$.
By proposition \ref{prop: concavity/convexity of vp}, $G'(m)\leq 0$ for all $m\in\R$ and equality holds iff ($J=J_c$ and $m=\phi_1(h,J_c)=\phi_2(h,J_c)\,$). Hence $G$ is strictly decreasing on $\R$.
On the other hand by (\ref{eq: vp incr m<=0, decr m>=1}), $G(m)<0$ for all $m\leq0$ and $G(m)>0$
for all $m\geq1$.
Therefore there exists a unique point $m$ ($m\in\,]0,1[\,$) such that $G(m)=0$.\\[2pt]
$\bullet$ Suppose $J>J_c$.
By proposition \ref{prop: concavity/convexity of vp}, %$G'(m)<0$ for $m<\phi_1(h,J)$ or $m>\phi_2(h,J)$ and $G'(m)>0$ for $\phi_1(h,J)<m<\phi_2(h,J)$. $G$ is strictly decreasing on $\,]-\infty,\phi_1(h,J)]$, strictly increasing on $[\phi_1(h,J),\phi_2(h,J)]\,$ and again strictly decreasing on $[\phi_2(h,J),\infty[\,$.
$G$ is strictly decreasing for $m\leq\phi_1(h,J)$, strictly increasing for $\phi_1(h,J)\leq m\leq\phi_2(h,J)$ and again strictly decreasing for $m\geq\phi_2(h,J)$.
On the other hand by (\ref{eq: vp incr m<=0, decr m>=1}), $G(m_+)>0$ for some point $m_+<\phi_1(h,J)$ and $G(m_-)>0$ for some point $m_->\phi_2(h,J)$.
Therefore:
\begin{gather*}
(\,\exists\,(\text{a$\!$ unique})\;m_1\in\,]-\infty,\phi_1(h,J)]\ \text{s.t.}\ G(m_1)=0\,)\
\Leftrightarrow\ G(\phi_1(h,J))\leq0 \,; \\[4pt]
(\,\exists\,(\text{a$\!$ unique})\;m_2\in[\phi_2(h,J),\infty[\,\ \text{s.t.}\ G(m_2)=0\,)\
\Leftrightarrow\ G(\phi_2(h,J))\geq0 \,; \\[4pt]
(\,\exists\,(\text{a$\!$ unique})\;m_0\in[\phi_1(h,J),\phi_2(h,J)]\,\ \text{s.t.}\ G(m_0)=0\,)\
\Leftrightarrow\ G(\phi_1(h,J))\leq0 \,,\; G(\phi_2(h,J))\geq0 \,.
\end{gather*}
And now, using identity (\ref{eq: deriv vp w.r.t. m}) and definitions (\ref{eq: phi(h,J)}), (\ref{eq: psi(J)})
\[ G(\phi_1(h,J)) \underset{(=)}{<} 0\ \Leftrightarrow\ g\big((2\phi_1(h,J)-1)J+h\big) \underset{(=)}{<} \phi_1(h,J)\
\Leftrightarrow\ h \underset{(=)}{<} \psi_1(J) \]
and similarly $G(\phi_2(h,J))\underset{(=)}{>}0\ \Leftrightarrow\ h\underset{(=)}{>}\psi_2(J)\,$.\\[2pt]
%\[\begin{split}
%G(\phi_1(h,J))\underset{(=)}{<}0\ &\Leftrightarrow\ g\big((2\phi_1(h,J)-1)J+h\big) \underset{(=)}{<} \phi_1(h,J) \\
%&\Leftrightarrow\ g\big(\frac{1}{2}\,\log a_1(J)\big) \underset{(=)}{<} \frac{1}{2}-\frac{h}{2J}+\frac{1}{4J}\,\log a_1(J) \\
%&\Leftrightarrow\ h\underset{(=)}{<}\psi_1(J) \;;
%\end{split}\]
%and similarly $G(\phi_2(h,J))\underset{(=)}{>}0\ \Leftrightarrow\ h\underset{(=)}{>}\psi_2(J)\,$.
%Therefore:
%\vspace{-2pt}
%\begin{gather*}
%\exists\,(\text{a unique})\;m_1\in\,]-\infty,\phi_1(h,J)]\ \text{s.t.}\ G(m_1)=0
%\quad\Leftrightarrow\ \ h\leq\psi_1(J) \;; \\[4pt]
%\exists\,(\text{a unique})\;m_2\in[\phi_2(h,J),\infty[\,\ \text{s.t.}\ G(m_2)=0
%\quad\Leftrightarrow\ \ h\geq\psi_2(J) \;; \\[4pt]
%\exists\,(\text{a unique})\;m_0\in[\phi_1(h,J),\phi_2(h,J)]\,\ \text{s.t.}\ G(m_0)=0
%\quad\Leftrightarrow\ \ \psi_2(J)\leq h\leq\psi_1(J)\;;
%\end{gather*}
%and $m_1=\phi_1(h,J)=m_0$ iff $h=\psi_1(J)$, while $m_2=\phi_2(h,J)=m_0$ iff $h=\psi_2(J)$. \\[2pt]
%
The first $\bullet$ allows to conclude in case \textit{1.}, while the second $\bullet$ allows to conclude in all the other cases.
Notice that the nature of the stationary points of $\vpim(\cdot\,,h,J)$ %(i.e. the zeros of $G=\frac{\partial\vpim}{\partial m}$)
is determined by the sign of the second derivative $\frac{\partial^2\vpim}{\partial m^2}$ studied in proposition \ref{prop: concavity/convexity of vp}.
\endproof

\begin{figure}[h]
\centering
\includegraphics{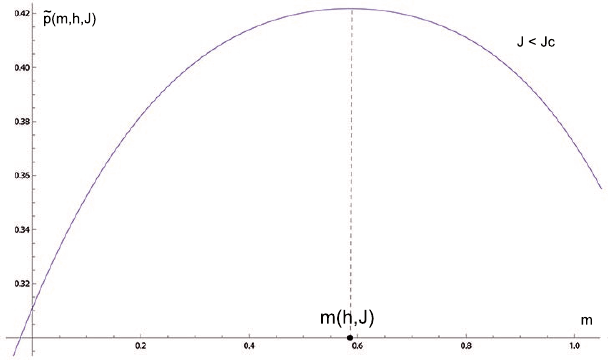} \
\includegraphics{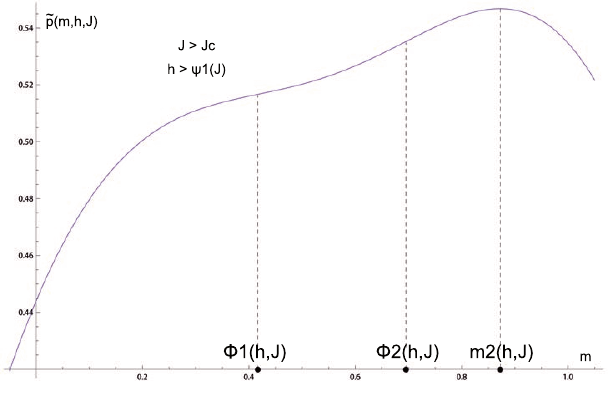} \\
\includegraphics{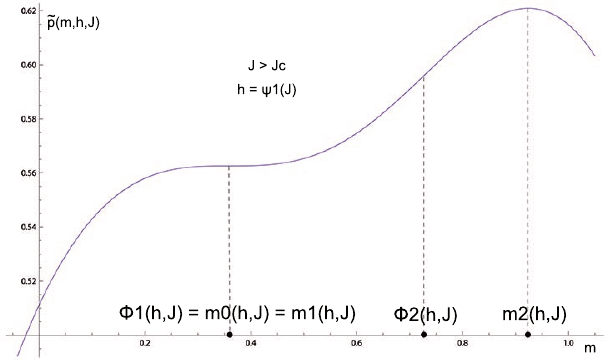} \
\includegraphics{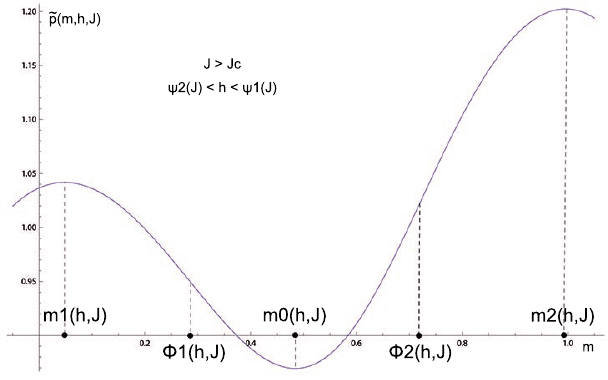}
\caption{Plots of the function $m\mapsto\vpim(m,h,J)$ for different values of the parameters $h,J$.
In particular cases \textit{1.}, \textit{3.}, \textit{4.}, \textit{2.} of proposition \ref{prop: stationary points of vp} are represented.}
\end{figure}

A special role is played by the point $(h_c,J_c)$, where we set
\begin{equation} \label{eq: hc}
h_c := \psi_1(J_c) = \psi_2(J_c) \,=\, \frac{1}{2}\,\log(2\sqrt{2}-2)-\frac{1}{4} \,\approx\, -0.3441 \;,
\end{equation}
indeed in the next sub-sections it will turn out to be the \textit{critical point} of the system.
It is also useful to define
\begin{gather}
m_c := \phi_1(h_c,J_c) = \phi_2(h_c,J_c) \,=\, 2-\sqrt{2} \,\approx\, 0.5857 \;, \label{eq: mc}\\
\xi_c := (2m_c-1)J_c+h_c \,=\, \frac{1}{2}\,\log(2\sqrt{2}-2) \,\approx\, -0.0941 \;. \label{eq: xic}
\end{gather}
The computations are done observing that $a_1(J_c)=a_2(J_c)=2\sqrt{2}-2\,$ and $g(\frac{1}{2}\log(2\sqrt{2}-2))=2-\sqrt{2}$.

\begin{rk} \label{rk: mc=m(hc,Jc)}
We notice that $m_c$ is the (unique) solution of equation (\ref{eq: ce}) for $h=h_c$ and $J=J_c$, that is $m(h_c,J_c)=m_c$.
Indeed a direct computation using (\ref{eq: g(h)}) shows
%\begin{equation}
%(2m_c-1)J_c + h_c \,=\, \frac{1}{2}\,\log(2\sqrt{2}-2)
%\end{equation}
%hence by (\ref{eq: g(h)})
\[\begin{split}
g\big((2m_c-1)J_c+h_c\big) \,&=\, g\big(\xi_c\big) \,=m_c.\\
\end{split}\]
%Furthermore $m_c$ is a solution of equation (\ref{eq: ce}) for all $h,J$ such that $h-h_c=-(2m_c-1)\,(J-J_c)$.
%Indeed
%\[\begin{split}
%g\big((2m_c-1)J+h\big) \,&=\, g\big((2m_c-1)J+h_c-(2m_c-1)(J-J_c)+\big) \,=\\
%&=\, g\big((2m_c-1)J_c+h_c\big) \,=\, m_c \;.
%\end{split}\]
Observe that as a consequence $m_c$ is a solution of equation (\ref{eq: ce}) for all $(h,J)$ such that $h-h_c=(1-2m_c)(J-J_c)$.
\end{rk}

In the next proposition we analyse the regularity of the solutions of equation (\ref{eq: ce}).

\begin{prop}[Regularity properties] \label{prop: continuity of stationary points of vp}
Consider the stationary points of $\vpim(\cdot\,,h,J)$ defined in proposition \ref{prop: stationary points of vp}: $m(h,J),\,m_1(h,J),\,m_0(h,J),\,m_2(h,J)$ for suitable values of $h,J$.
The functions
\begin{gather}
\mu_1(h,J):=\, \begin{cases}
\,m(h,J) & \text{if }\ 0<J\leq J_c\,,\ h\in\R \\
\,m_1(h,J) & \text{if }\ J>J_c\,,\ h\leq\psi_1(J)
\end{cases}\;; \\[2pt]
\mu_2(h,J):=\, \begin{cases}
\,m(h,J) & \text{if }\ 0<J\leq J_c\,,\ h\in\R \\
\,m_2(h,J) & \text{if }\ J>J_c\,,\ h\geq\psi_2(J)
\end{cases}\;; \\[2pt]
\mu_0(h,J):=\, \begin{cases}
\,m(h,J) & \text{if }\ 0<J\leq J_c\,,\ h\in\R \\
\,m_0(h,J) & \text{if }\ J>J_c\,,\ \psi_2(J)\leq h\leq\psi_1(J)
\end{cases}\;.
\end{gather}
have the following properties:
\begin{itemize}
\item[i)] are continuous on the respective domains;
\item[ii)] are $C^\infty$ in the interior of the respective domains;
\item[iii)] for $i=0,1,2$ and $(h,J)$ in the interior of the domain of $\mu_i$
\begin{gather}
\frac{\partial}{\partial h}\,\vpim(\mu_i(h,J),h,J) \,=\, \mu_i \;,\quad
\frac{\partial}{\partial J}\,\vpim(\mu_i(h,J),h,J) \,= -\,\mu_i\,(1-\mu_i) \;; \label{eq: deriv vp(m) w.r.t. h, J} \\[4pt]
\frac{\partial\mu_i}{\partial h} \,=\, \frac{2\,\mu_i\,(1-\mu_i)}{2-\mu_i-4J\,\mu_i\,(1-\mu_i)} \;,\qquad
\frac{\partial\mu_i}{\partial J} \,=\, (2\mu_i-1)\,\frac{\partial\mu_i}{\partial h} \;. \label{eq: deriv m w.r.t. h, J}
\end{gather}
\end{itemize}
\end{prop}

\proof
\textit{i)} First prove the continuity of $\mu_1$.
%Remember that by remark \ref{rk: extremum iff ce} they take values in  $]0,1[$.\\
Observe that by propositions \ref{prop: stationary points of vp}, \ref{prop: concavity/convexity of vp}:
\begin{itemize}
\item for $(h,J)$ in $D_1:=\{(h,J) \,|\, (0<J\leq J_c,\,h\in\R)\;\text{or}\;(J>J_c,\,h\leq\psi_2(J))\}\,$, $\mu_1(h,J)$ is the \textit{only} maximum point of $\vpim(\cdot\,,h,J)$ on the interval $[0,1]\,$;
\item for $(h,J)$ in $D_2:=\,\{(h,J) \,|\, J\geq J_c,\,h\leq\psi_1(J)\}\,$, $\mu_1(h,J)$ is the \textit{only} maximum point of $\vpim(\cdot\,,h,J)$ on the interval $[0,\phi_1(h,J)]\,$.
\end{itemize}
Hence by proposition \ref{propA: continuity of the max, argmax}, continuity of the functions $\vpim$ and $\phi_1$ implies continuity of the function $\mu_1$ on the sets $D_1$ and $D_2$.
%\begin{gather*}
%D_1 :=\, \{(h,J) \;|\; (0<J\leq J_c,\, h\in\R)\ \text{or}\ (J>J_c,\ h\leq\psi_2(J)) \} \;,\\
%D_2 :=\, \{(h,J) \;|\; J\geq J_c,\, h\leq\psi_1(J)) \} \;.
%\end{gather*}
As $D_1$ and $D_2$ are both closed subsets of $\R\times\R_+$, by the pasting lemma $\mu_1$ is continuous on their union
\[ D_1\cup D_2 \,=\, \{(h,J) \;|\; (0<J\leq J_c,\, h\in\R)\ \text{or}\ (J>J_c,\, h\leq\psi_1(J)) \} \;.\]
A similar argument proves the continuity of $\mu_2$ and $\mu_0$.\\[2pt]
\textit{ii)} Now prove the smoothness of $\mu_1,\,\mu_2,\,\mu_0$ in the \textit{interior} of their domains.
Set $G(m,h,J):=\frac{\partial\vpim}{\partial m}(m,h,J)$. As just seen $m=\mu_1(h,J),\,\mu_2(h,J),\,\mu_0(h,J)$ are \textit{continuous} solutions of
\[G(m,h,J)=0\;,\]
for values of $h,J$ in the respective domains.
Observe that $G\in C^\infty(\R\times\R\times\R_+)$ and by propositions \ref{prop: concavity/convexity of vp}, \ref{prop: stationary points of vp} it can happen
\[\begin{split}
&\begin{cases}
\dfrac{\partial G}{\partial m}\,(m,h,J) = 0 \\[2pt]
G(m,h,J) = 0
\end{cases} \Leftrightarrow\ \begin{cases}
J\geq J_c\,,\;(m=\phi_1(h,J)\text{ or }m=\phi_2(h,J)) \\[2pt]
G(m,h,J) = 0
\end{cases} \!\!\Leftrightarrow\\[4pt]
&\Leftrightarrow\ \begin{cases}
J\geq J_c,\, m=\phi_1(h,J)\\[2pt]
h=\psi_1(J)
\end{cases} \!\!\text{or }\ \begin{cases}
J\geq J_c,\, m=\phi_2(h,J)\\[2pt]
h=\psi_2(J)
\end{cases} \!\!.
\end{split}\]
%&\Leftrightarrow\ J\geq J_c\,,\;(m=\phi_1(\psi_1(J),J)\text{ or }m=\phi_2(\psi_2(J),J))
$m=\mu_1(h,J)$ can fall only within the first case, while $m=\mu_2(h,J)$ can fall only within the second case. Therefore by the implicit function theorem (corollary \ref{corA: regularity of a continuous solution}), $\mu_1,\,\mu_2,\,\mu_0$ are $C^\infty$ on the interior of the respective domains.\\[2pt]%respectively on $\textrm{Dom}(\mu_1)\,\smallsetminus\,\{(h,J)\,|\,J\geq J_c,\,h=\psi_1(J)\}$, $\textrm{Dom}(\mu_2)\smallsetminus\{(h,J)\,|\,J\geq J_c,\,h=\psi_2(J)\}\,\big)$, $\textrm{Dom}(\mu_0)\smallsetminus\{(h,J)\,|\,J\geq J_c,\,(h=\psi_1(J)\text{ or }h=\psi_1(J))\}$.\\[2pt]
\textit{iii)} Let $i=0,1,2$ and $(h,J)$ in the interior of the domain of $\mu_i$.
Using (\ref{eq: vp(m,h,J)}), $(\pcl)'=g$ and the fact that $\mu_i(h,J)$ satisfies equation (\ref{eq: ce}), compute
\[\begin{split}
\frac{\partial}{\partial h}\,\vpim(\mu_i,h,J) \,&=
-2J\,\frac{\partial\mu_i}{\partial h} + (\pcl)'\big((2\mu_i-1)J+h\big)\, (2J\,\frac{\partial\mu_i}{\partial h}+1) \\
&= -2J\,\frac{\partial\mu_i}{\partial h} + \mu_i\, (2J\,\frac{\partial\mu_i}{\partial h}+1) \,=\,
\mu_i \;;
\end{split}\]
%\begin{split}
%\frac{\partial}{\partial J}\,\vpim(\mu_i,h,J) \,&=
%-\mu_i^2 -2J\,\mu_i\,\frac{\partial\mu_i}{\partial h} + (\pcl)'\big((2\mu_i-1)J+h\big)\,(2\mu_i-1 + 2J\,\frac{\partial\mu_i}{\partial J}\,) \\
%&=-\mu_i^2 -2J\,\mu_i\,\frac{\partial\mu_i}{\partial h} + \mu_i\,(2\mu_i-1 + 2J\,\frac{\partial\mu_i}{\partial J}) \,=\,
%\mu_i^2-\mu_i \;.
%\end{split}
and similarly $\frac{\partial}{\partial J}\,\vpim(\mu_i,h,J) = \mu_i^2-\mu_i\,$.\\[2pt]
Using the fact that $\mu_i(h,J)$ satisfies equation (\ref{eq: ce}) compute
\[\begin{split}
&\frac{\partial\mu_i}{\partial h} \,=\, \frac{\partial}{\partial h}\,g\big((2\mu_i-1)J+h\big) \,=\,
g'\big((2\mu_i-1)J+h)\big)\,(1+2J\,\frac{\partial\mu_i}{\partial h}) \\
&\Rightarrow\ \frac{\partial\mu_i}{\partial h} \,=\, \frac{g'\big((2\mu_i-1)J+h\big)}{1-2J\,g'\big((2\mu_i-1)J+h\big)} \;;
\end{split}\]
%\begin{split}
%\frac{\partial\mu_i}{\partial J} \,=\, \frac{\partial}{\partial J}\,g\big((2\mu_i-1)J+h\big) \,=\,
%g'\big((2\mu_i-1)J+h)\big)\,(2\mu_i-1+2J\,\frac{\partial\mu_i}{\partial J}) \\
%&\Rightarrow\ \frac{\partial\mu_i}{\partial J} \,=\, \frac{(2\mu_i-1)\,g'\big((2\mu_i-1)J+h\big)}{1-2J\,g'\big((2\mu_i-1)J+h\big)} \;;
%\end{split}
and similarly $\frac{\partial\mu_i}{\partial J} = \frac{(2\mu_i-1)\,g'\big((2\mu_i-1)J+h\big)}{1-2J\,g'\big((2\mu_i-1)J+h\big)}\,$.
Then observe that $g'=2\,g\,(1-g)/(2-g)$ (identity (\ref{eq: g' function of g}) in the Appendix), hence since $\mu_i(h,J)$ satisfies equation (\ref{eq: ce})
\[ g'\big((2\mu_i-1)J+h\big) \,=\, \frac{2\,\mu_i\,(1-\mu_i)}{2-\mu_i} \;;\]
substituting this in the previous identities concludes the proof.
\endproof

To end this subsection we study the asymptotic behaviour of the stationary points of $\vpim(\cdot\,,h,J)$ for large $J$.

\begin{prop}[Asymptotic behaviour] \label{prop: m for J->infty}
Consider the stationary points $m_1(h,J)$, $m_0(h,J)$, $m_2(h,J)$ defined in proposition \ref{prop: stationary points of vp} for suitable values of $h,J$.
\begin{itemize}
\item[i)] For all fixed $h\in\R$
\[ m_1(h,J)\,\xrightarrow[J\to\infty]{}\,0\;,\quad
m_2(h,J)\,\xrightarrow[J\to\infty]{}\,1\;,\quad
m_0(h,J)\,\xrightarrow[J\to\infty]{}\,\frac{1}{2} \;.\]
\item[ii)] Moreover for all fixed $h\in\R$
\[ J\,m_1(h,J)\,\xrightarrow[J\to\infty]{}\,0 \;,\quad
J\,(1-m_2(h,J))\,\xrightarrow[J\to\infty]{}\,0 \;.\]
\item[iii)] And taking the $\sup$ and $\inf$ over $h\in[\psi_2(J),\psi_1(J)]\,$
\[\sup_{h} m_1(h,J)\,\xrightarrow[J\to\infty]{}\,0 \;,\quad
\inf_{h} m_2(h,J)\,\xrightarrow[J\to\infty]{}\,1 \;.\]
\end{itemize}
\end{prop}

\proof
\textit{i)} First observe from the definition (\ref{eq: psi(J)}) that $\psi_2(J)\rightarrow-\infty$, $\psi_1(J)\rightarrow\infty$ as $J\to\infty$. Hence for any fixed $h\in\R$ there exists $\bar J>0$ such that $\psi_2(J)<h<\psi_1(J)$ for all $J>\bar J$. This means that the limits in the statement make sense.\\
Now remind that by proposition \ref{prop: stationary points of vp}, for $J>\bar J$
\[ m_1(h,J) < \phi_1(h,J) < m_0(h,J) < \phi_2(h,J) < m_2(h,J) \;.\]
Observe from the definition (\ref{eq: phi(h,J)}) that $\phi_1(h,J)\rightarrow\frac{1}{2}\,$, $\phi_2(h,J)\rightarrow\frac{1}{2}\,$ as $J\to\infty$. It follows immediately that also $m_0(h,J) \rightarrow \frac{1}{2}$ as $J\to\infty$.\\
Moreover definition (\ref{eq: phi(h,J)}) entails that $J\big(\frac{1}{2}-\phi_1(h,J)\big)\rightarrow\infty\,$, $J\big(\phi_2(h,J)-\frac{1}{2}\big)\rightarrow\infty$ as $J\to\infty$.
Exploit the fact that $m_1(h,J)$ is a solution of equation (\ref{eq: ce}):
\[\begin{split}
m_1(h,J) \,&=\, g\big((2m_1(h,J)-1)\,J+h\big) \,\leq\, g\big((2\phi_1(h,J)-1)\,J+h\big) \,=\\
&=\, g\big(-2J\,(\frac{1}{2}-\phi_1(h,J))+h\big) \,\xrightarrow[J\to\infty]{}\, 0 \;,
\end{split}\]
where also the facts that the function $g$ is increasing and $g(\xi)\rightarrow0$ as $\xi\to-\infty$ are used.
As by remark \ref{rk: extremum iff ce} $m_1$ takes values in $]0,1[$, conclude that $m_1(h,J)\longrightarrow0$ as $J\to\infty$.
Similarly it can be shown that $m_2(h,J)\longrightarrow1$ as $J\to\infty$.\\[2pt]
%Similarly exploit the fact that $m_2(h,J)$ is a solution of equation (\ref{eq: ce}):
%\[\begin{split}
%m_2(h,J) \,&=\, g\big((2m_2(h,J)-1)\,J+h\big) \,\geq\, g\big((2\phi_2(h,J)-1)\,J+h\big) \,=\\
%&=\, g\big(2J\,(\phi_2(h,J)-\frac{1}{2})+h\big) \,\xrightarrow[J\to\infty]{}\, 1 \;,
%\end{split}\]
%where also the facts that the function $g$ is increasing and $g(\xi)\rightarrow1$ as $\xi\to\infty$ are used.
%As by remark \ref{rk: extremum iff ce} $m_1$, $m_2$ take values in $]0,1[$, conclude that $m_1(h,J)\rightarrow0\,$, $m_2(h,J)\rightarrow1$ as $J\to\infty$.\\[2pt]
%
\textit{ii)} Start observing that, by a standard computation from the definition (\ref{eq: g(h)}), $\xi\,g(-\xi)\longrightarrow0$ and $\xi\,\big(1-g(\xi)\big)\longrightarrow0$ as $\xi\to+\infty$.
Then exploit the fact that, for fixed $h$ and $J$ sufficiently large, $m_1=m_1(h,J)$ %, $m_2=m_2(h,J)$
is a solution of equation (\ref{eq: ce}):
\[\begin{split}
&J\,m_1 \,=\, J\,g\big((2m_1-1)J+h\big) \,=\\[2pt]
&= \frac{\big((1-2m_1)J-h\big)\;g\big(\!-(1-2m_1)J+h\big)}{1-2m_1}\, + \frac{h\;g\big(\!-(1-2m_1)J+h\big)}{1-2m_1}
\,\xrightarrow[J\to\infty]{}\, \frac{0}{1}\, + \frac{h\,0}{1} \,= 0 \,,
\end{split}\]
using also that $m_1\rightarrow 0$ as $J\to\infty$ by \textit{i)}.
Similarly it can be shown that $J\,(1-m_2)\longrightarrow0$ as  $J\to\infty$.\\[2pt]
%\[\begin{split}
%&J\,(1-m_2) \,=\, J\,\big(\,1-g\big((2m_2-1)J+h\big)\,\big) \,=\\[2pt]
%&= \frac{\big((2m_2-1)J+h\big)\;\big(\,1-g\big((2m_2-1)J+h\big)\,\big)}{2m_2-1}\, - \frac{h\,\big(\,1-g\big((2m_2-1)J+h\big)\,\big)}{2m_2-1}
%\,\xrightarrow[J\to\infty]{}\, \frac{0}{1}\, + \frac{h\,0}{1} \,= 0 \,,
%\end{split}\]
%because $m_2\longrightarrow 1$ as $J\to\infty$ by \textit{i)}.\\[2pt]
%
\textit{iii)} Start observing that, by a standard computation from the definition (\ref{eq: psi(J)}), $-J+\psi_1(J)\longrightarrow-\infty$ and $J+\psi_2(J)\longrightarrow\infty$ as $J\to\infty$.
Then exploit the fact that, for $J>J_c$ and $h\in[\psi_2(J),\psi_1(J)]$, $m_1=m_1(h,J)$ %, $m_2=m_2(h,J)$
is a solution of equation (\ref{eq: ce}):
\[\begin{split}
\sup_{h\in[\psi_2,\psi_1]}m_1 \,&=\, \sup_{h\in[\psi_2,\psi_1]}g\big((2m_1-1)J+h\big) \,\leq\, g\big((2m_1-1)J+\psi_1(J)\big) \,=\\ &=\, g\big(2J\,m_1-J+\psi_1(J)\big) \,\xrightarrow[J\to\infty]{}\,0 \;,
\end{split}\]
using also the facts that $g$ is an increasing function, $g(\xi)\to0$ as $\xi\to-\infty$, and $J\,m_1\rightarrow0$ as $J\to\infty$ by \textit{ii)}.
Similarly it can be shown that $\inf_{h\in[\psi_2,\psi_1]}m_2\longrightarrow1$ as $J\to\infty$.
%\[\begin{split}
%\inf_{h\in[\psi_2,\psi_1]}m_2 \,&=\, \inf_{h\in[\psi_2,\psi_1]}g\big((2m_2-1)J+h\big) \,\geq\, g\big((2m_2-1)J+\psi_2(J)\big) \,=\\ &=\, g\big(\!-2J\,(1-m_2)+J+\psi_2(J)\big) \,\xrightarrow[J\to\infty]{}\,1 \;,
%\end{split}\]
%where also the facts that $g$ is an increasing function, $g(\xi)\to1$ as $\xi\to\infty$ and $J\,(1-m_2)\longrightarrow0$ as $J\to\infty$ by \textit{ii)}.
\endproof
\subsection{The ``wall'': existence and uniqueness, regularity and asymptotic behavior }
In the previous subsection we studied all the solutions of equation (\ref{eq: ce}), that is all the stationary points of $m\mapsto\vpim(m,h,J)$. One of them is the point where the global maximum is attained and, because of theorem \ref{thm: imitative mean field limit} and remark \ref{rk: m*=monomer density}, we are interested in this one.\\
Consider the points $m,\,m_1,\,m_0,\,m_2$ defined in proposition \ref{prop: stationary points of vp} and look for the global maximum point of $m\mapsto\vpim(m,h,J)$:
\begin{itemize}
\item for $0<J<J_c$ and $h\in\R$, $m(h,J)$ is the only local maximum point, hence it is the global maximum point;
\item for $J>J_c$ and $h\leq\psi_2(J)$, $m_1(h,J)$ is the only local maximum point, hence it is the global maximum point;
\item for $J>J_c$ and $h\geq\psi_1(J)$, $m_2(h,J)$ is the only local maximum point, hence it is the global maximum point;
\item for $J>J_c$ and $\psi_2(J)<h<\psi_1(J)$, there are two local maximum points $m_1(h,J) < m_2(h,J)$, hence at least one of them is the global maximum point.
\end{itemize}
To answer which one is the global maximum point in the last case, we have to investigate the sign of the following function
\begin{equation} \label{eq: Delta(h,J)}
\Delta(h,J) :=\, \vpim\big(m_2(h,J),h,J\big)-\vpim\big(m_1(h,J),h,J\big)
\end{equation}
for $J>J_c$ and $\psi_2(J)\leq h\leq\psi_1(J)\,$.

\begin{prop}[Existence and Uniqueness] \label{prop: wall existence uniqueness}
 For all $J>J_c$ there exists a unique $h=\gamma(J)\in\,]\psi_2(J),\psi_1(J)[\,$ such that $\Delta(h,J)=0$.
Moreover
\[ \Delta(h,J) \begin{cases}
<0 & \text{if }\ J>J_c,\ \psi_2(J)\leq h<\gamma(J) \\[2pt]
>0 & \text{if }\ J>J_c,\ \gamma(J)<h\leq\psi_1(J)
\end{cases} \ .\]
\end{prop}

\proof
It is an application of the intermediate value theorem. Fix $J>J_c$. It suffices to observe that
\begin{itemize}
\item[i.] $\Delta\big(\psi_2(J),J\big)<0$, because for $h=\psi_2(J)$ the only maximum point of the function $\vpim(\cdot\,,h,J)$ is $m_1(h,J)$;
\item[ii.] $\Delta\big(\psi_1(J),J\big)>0$, because for $h=\psi_1(J)$ the only maximum point of the function $\vpim(\cdot\,,h,J)$ is $m_2(h,J)$;
\item[iii.] $h\mapsto\Delta(h,J)$ is a continuous function, by continuity of $\vpim$, $m_1$, $m_2$ (see proposition \ref{prop: continuity of stationary points of vp});
\item[iv.] $h\mapsto\Delta(h,J)$ is strictly increasing; indeed it is $C^\infty$ on $\,]\psi_2(J),\psi_1(J)[\,$ by smoothness of $\vpim$, $m_1$, $m_2$ (see proposition \ref{prop: continuity of stationary points of vp}) and, by formula (\ref{eq: deriv vp(m) w.r.t. h, J}),
%with a computation analogous to that one done in remark \ref{rk: m*=monomer density} one finds
\[\begin{split}
\frac{\partial\Delta}{\partial h}\,(h,J) \,&=\,
\frac{\partial}{\partial h}\,\vpim\big(m_2(h,J),h,J\big) - \frac{\partial}{\partial h}\,\vpim\big(m_1(h,J),h,J\big) \,=\\
&=\, m_2(h,J) - m_1(h,J) \,>\, \phi_2(h,J) - \phi_1(h,J) \,>\, 0
\end{split}\]
for all $h\in\,]\psi_2(J),\psi_1(J)[\,$.\qedhere
\end{itemize}
\endproof

\begin{figure}[h]
\centering
\includegraphics{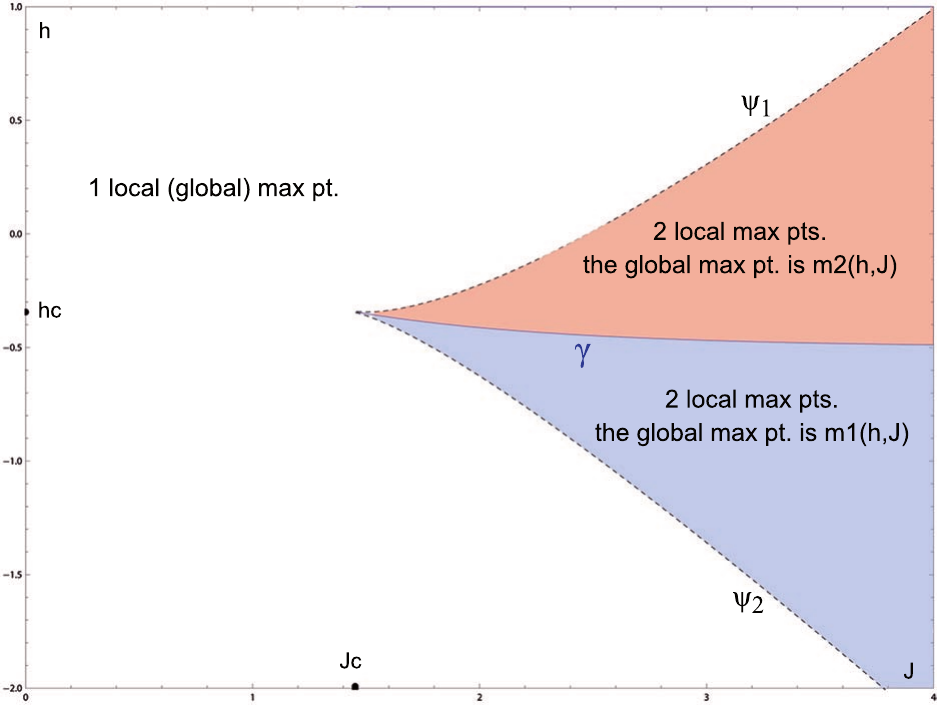}
\caption{$\gamma$ separates the values of $h,J$ for which $m_1(h,J)$ is the global maximum point  from those for which $m_2(h,J)$ is the global maximum point of $m\mapsto\vpim(m,h,J)$.
As $m_1(h,J)<m_2(h,J)$, this entails a discontinuity of the global maximum point $m^*(h,J)$ along the ``\textit{wall}''  $\Gamma$.}
\end{figure}

\begin{rk} \label{rk: global max m*}
By the previous results the global maximum point of $m\mapsto\vpim(m,h,J)$ is
\begin{equation}
m^*(h,J) := \begin{cases}
\,m(h,J) & \text{if }\; 0<J\leq J_c\,,\ h\in\R \\
\,m_1(h,J) & \text{if }\; J>J_c\,,\ h<\gamma(J) \\
\,m_2(h,J) & \text{if }\; J>J_c\,,\ h>\gamma(J)
\end{cases}
\end{equation}
where the function $\gamma$ is defined by proposition \ref{prop: wall existence uniqueness}.
Set also
\begin{equation}
\Gamma := \{(h,J) \;|\; J>J_c,\; h=\gamma(J) \} \;,\qquad
\overline\Gamma :=\, \Gamma \,\cup\, \{(h_c,J_c)\}\ .
\end{equation}
Notice that proposition \ref{prop: wall existence uniqueness} guarantees that there is only a curve $\Gamma$ in the plane $(h,J)$ where the global maximum point of $m\mapsto\vpim(m,h,J)$ is not unique. We leaved the function $m^*$ undefined on $\Gamma$.\\
By proposition \ref{prop: continuity of stationary points of vp} it follows that the function $m^*$ is continuous on its domain $(\R\times\R_+)\smallsetminus\Gamma$ and it is $C^\infty$ on $(\R\times\R_+)\smallsetminus\overline\Gamma\,$.
The behaviour of $m^*$ at the critical point $(h_c,J_c)$ will be investigated in the next subsection.
\end{rk}

Now we investigate the main properties of the curve $\overline\Gamma$, which we call ``\textit{the wall}''.
Extend the function $\gamma$ defined by proposition \ref{prop: wall existence uniqueness} by
\begin{equation} \label{eq: gamma}
\overline\gamma(J) := \begin{cases}
\,\gamma(J) & \text{if }\; J>J_c \\
\,h_c & \text{if }\; J=J_c
\end{cases}\ .
\end{equation}

\begin{prop}[Regularity properties] \label{prop: wall regularity}
The function $\overline\gamma$ is $C^\infty$ on $]J_c,\infty[\,$ and (at least) $C^1$ on $[J_c,\infty[$.
In particular
\[ \gamma'(J) \,=\, 1- m_1\big(\gamma(J),J\big) - m_2\big(\gamma(J),J\big) \quad\forall\,J>J_c \;,\]
and
\[ \overline\gamma\,'(J_c) \,=\, 1-2\,m_c \,=\, -(3-2\sqrt{2}) \;.\]
\end{prop}

\proof
\textit{I.} First prove that the function $\gamma\in C^\infty(\,]J_c,\infty[\,)$.\\
By proposition \ref{prop: wall existence uniqueness} for all $J>J_c$, $h=\gamma(J)$ is the \textit{unique} solution of equation
\[ \Delta(h,J) = 0 \]
where $\Delta$ is defined by (\ref{eq: Delta(h,J)}). Moreover $\psi_2(J)<\gamma(J)<\psi_1(J)$.
Observe that $\Delta$ is $C^\infty$ on $\{(h,J)\,|\,J>J_c,\;\psi_2(J)<h<\psi_1(J)\}$ by smoothness of $\vpim$ and $m_1$, $m_2$ on this region (see proposition \ref{prop: continuity of stationary points of vp}).
And furthermore, as shown in the proof of proposition \ref{prop: wall existence uniqueness},
\[ \frac{\partial\Delta}{\partial h}\,(h,J) \neq 0 \quad\forall\,(h,J)\text{ s.t. }h=\gamma(J) \,.\]
Therefore by the implicit function theorem (corollary \ref{corA: regularity of a unique solution}), $\gamma\in C^\infty(\,]J_c,\infty[\,)$. Now
\[\begin{split}
&\Delta(\gamma(J),J))\equiv0 \ \Rightarrow\
0 = \frac{\dd}{\dd J}\,\Delta(\gamma(J),J) = \frac{\partial\Delta}{\partial h}(\gamma(J),J)\,\gamma'(J) + \frac{\partial\Delta}{\partial J}(\gamma(J),J) \\
&\Rightarrow\ \gamma'(J) \,= -\,\frac{\partial\Delta}{\partial J}\,\big/\,\frac{\partial\Delta}{\partial h}\;(\gamma(J),J)\;;
\end{split}\]
by formulae (\ref{eq: deriv vp(m) w.r.t. h, J}) $\frac{\partial\Delta}{\partial h} = m_2-m_1$ and $\frac{\partial\Delta}{\partial J} = (m_2^2-m_2)-(m_1^2-m_1)\,$;
%\begin{gather*}
%\frac{\partial\Delta}{\partial h}\,(h,J) \,=\, (m_2-m_1)\,(h,J) \;,\\[2pt]
%\frac{\partial\Delta}{\partial J}\,(h,J) \,=\, \big((m_2^2-m_2)-(m_1^2-m_1)\big)\,(h,J) \;;
%\end{gather*}
therefore
\[\begin{split}
\gamma'(J) \,&= \, 1-(m_2+m_1)\,(\gamma(J),J) \;.
\end{split}\]
\textit{II.} Now prove that the extended function $\overline\gamma\in C^1([J_c,\infty[)\,$.\\
First observe that $\overline\gamma$ is continuous also in $J_c$, indeed:
\[ \psi_2(J)<\gamma(J)<\psi_1(J)\ \ \forall\,J>J_c \quad\Rightarrow\quad \lim_{J\to J_c+}\gamma(J)=h_c  \]
by definition of $h_c$ (\ref{eq: hc}) and continuity of $\psi_1,\,\psi_2$.
Then observe that
\[ \gamma'(J) \,=\, 1-(m_2+m_1)\,(\gamma(J),J)\ \xrightarrow[J\to J_c+]{}\ 1-2m_c \]
because $m(h_c,J_c)=m_c$ (remark \ref{rk: mc=m(hc,Jc)}) and the functions $\mu_1,\,\mu_2$ defined in proposition \ref{prop: continuity of stationary points of vp} are continuous.
%the elementary lemma \ref{lemA: elementary property of derivative},
By an immediate application of the mean value theorem, this proves that there exists $\overline\gamma\,'(J_c) = 1-2m_c$.
%and so $\overline\gamma\in C^1([J_c,\infty[)$.
%Using $\mu_1,\,\mu_2$ one may write
%\[ \overline\gamma\,'(J) \,=\, 1-(\mu_2+\mu_1)\,(\overline\gamma(J),J) \quad\forall\,J\geq J_c \]
\endproof

%The asymptotic behavior of the ``\textit{the wall}'' is described by the following:

\begin{prop}[Asymptotic behavior] \label{prop: wall asymptote}
 The function $\overline\gamma$ has an asymptote, precisely
\[ \gamma(J) \,\xrightarrow[J\to\infty]{}\, -\frac{1}{2} \;.\]
\end{prop}

\proof
\textit{I.} Consider the function $\Delta$ defined by (\ref{eq: Delta(h,J)}). The first step is to prove that $\Delta(h,J)\longrightarrow0$ as $J\to\infty$, $h=-\frac{1}{2}\,$.
Use identities (\ref{eq: vp(m,h,J)}), (\ref{eq: p(h)}) and the fact that for fixed $h$ and $J$ sufficiently large $m_1=m_1(h,J)$, $m_2=m_2(h,J)$ satisfy equation (\ref{eq: ce}), in two different ways:
%\[\begin{split}
%&\vpim(m_1,h,J) \,= -J\,m_1^2 +\frac{J}{2}\, + \pcl\big((2m_1-1)J+h\big) \,=\\
%&= -J\,m_1^2 +\frac{J}{2}\, -\frac{1-g\big((2m_1-1)J+h\big)}{2}\, - \log g\big((2m_1-1)J+h\big) + (2m_1-1)J+h \\
%&\phantom{\vpim(m_1,h,J)} \,= -J\,m_1^2 +\frac{J}{2}\, -\frac{1-m_1}{2}\, - \log g\big((2m_1-1)J+h\big) + (2m_1-1)J+h  \;;\\[4pt]
%&\vpim(m_2,h,J) \,= -J\,m_2^2 +\frac{J}{2}\, + \pcl\big((2m_2-1)J+h\big) \,=\\
%&= -J\,m_2^2 +\frac{J}{2}\, -\frac{1-g\big((2m_2-1)J+h\big)}{2}\, - \log g\big((2m_2-1)J+h\big) + (2m_2-1)J+h \\
%&\phantom{\vpim(m_2,h,J)} \,= -J\,m_2^2 +\frac{J}{2}\, -\frac{1-m_2}{2}\, - \log m_2 + (2m_2-1)J+h  \;.
%\end{split}\]
\begin{gather*}
\vpim(m_1,h,J) \,=\, -J\,m_1^2 +\frac{J}{2} -\frac{1-m_1}{2} - \log g\big((2m_1-1)J+h\big) + (2m_1-1)J+h \;,\\
\vpim(m_2,h,J) \,=\, -J\,m_2^2 +\frac{J}{2} -\frac{1-m_2}{2} - \log m_2 + (2m_2-1)J+h \;.
\end{gather*}
Hence, reminding that $m_1\rightarrow0$ and $m_2\rightarrow1$ as $J\to\infty$ by proposition \ref{prop: m for J->infty} part \textit{i)},
\[\begin{split}
\Delta(h,J) \,&=\, \vpim(m_2,h,J) - \vpim(m_1,h,J) \,=\\
&=\, J\,(-m_2^2+2m_2+m_1^2-2m_1) + \log g\big((2m_1-1)J+h\big) + \frac{1}{2} + o(1) \;,
\end{split}\]
%&=\, J\,(-m_2^2+2m_2+m_1^2-2m_1) + \frac{m_2-m_1}{2} - \log m_2 + \log g\big((2m_1-1)J+h\big) \\
Set $\delta:=-m_2^2+2m_2+m_1^2-2m_1$ and $\xi:=(2m_1-1)J+h$ and prove that in general
\begin{equation} \label{eq: wall asymptote 1}
J\,\delta + \log  g(\xi) \,\xrightarrow[J\to\infty]{}\, h \;;
\end{equation}
in particular it will follow that for $h=-\frac{1}{2}$
\begin{equation} \label{eq: wall asymptote 2}
\Delta\big(-\frac{1}{2},\,J\big) \,\xrightarrow[J\to\infty]{}\, 0 \;.
\end{equation}
Now proving (\ref{eq: wall asymptote 1}) is equivalent to prove $\exp(J\delta)\,g(\xi)\longrightarrow\exp(h)$ as $J\to\infty$;  and using definition (\ref{eq: g(h)})
\[ e^{J\delta}\,g(\xi) \,=\, e^{J\delta}\,\frac{\sqrt{e^{4\xi}+4e^{2\xi}}-e^{2\xi}}{2} \;=\,
\frac{\sqrt{e^{2(J\delta+2\xi)}+4e^{2(J\delta+\xi)}}-e^{J\delta+2\xi}}{2}\, \xrightarrow[J\to\infty]{}\, e^h \;,\]
%\frac{\sqrt{0+4e^{2h}}-0}{2} \,=\,
because, since $J\,m_1\rightarrow0$ and $J\,(1-m_2)\rightarrow0$ as $J\to\infty$ by proposition \ref{prop: m for J->infty} part \textit{ii)},
\begin{gather*}
J\delta+2\xi \,=\, J\,\big(\!-(1-m_2)^2+m_1^2-2m_1-1\,\big) + 2h \;\xrightarrow[J\to\infty]{}\, -\infty \;,\\[2pt]
J\delta+\xi \,=\, J\,\big(\!-(1-m_2)^2+m_1^2\,\big) + h \;\xrightarrow[J\to\infty]{}\; h \;.
\end{gather*}
\textit{II.} Remember that by definition of $\gamma$ in proposition \ref{prop: wall existence uniqueness}
\begin{equation} \label{eq: wall asymptote 3}
\Delta\big(\gamma(J),J\big) \,=\, 0 \quad\forall\,J>J_c \;;
\end{equation}
hence using (\ref{eq: wall asymptote 2}) will not be hard to prove that $\gamma(J)\longrightarrow-\frac{1}{2}$ as $J\to\infty$.
Let $\epsilon>0$. By (\ref{eq: wall asymptote 2}) there exists $\bar J_\epsilon>J_c$ such that
\begin{equation} \label{eq: wall asymptote 4}
\big|\Delta\big(\!-\frac{1}{2},\,J\big)\big| <\, \epsilon \quad\forall\,J>\bar J_\epsilon \;.
\end{equation}
Now by the mean value theorem for all $J>J_c$ and $h\in[\psi_2(J),\psi_1(J)]$,
\[ \big|\Delta(h,J)-\Delta\big(\!-\frac{1}{2},\,J\big)\big| \;\geq\;
\inf_{[\psi_2(J),\psi_1(J)]}\big|\frac{\partial\Delta}{\partial h}\,(\cdot\,,J)\big|\; \big|h+\frac{1}{2}\,\big| \;.\]
Furthermore by identity (\ref{eq: deriv vp(m) w.r.t. h, J}) and proposition \ref{prop: m for J->infty} part \textit{iii)}
\[\begin{split}
&\inf_{[\psi_2(J),\psi_1(J)]}\big|\frac{\partial\Delta}{\partial h}\,(\cdot\,,J)\big| \;=\,
\inf_{[\psi_2(J),\psi_1(J)]}(m_2-m_1)\,(\cdot\,,J) \,\geq\\[2pt]
&\geq\, \inf_{[\psi_2(J),\psi_1(J)]}m_2(\cdot,J) \;- \sup_{[\psi_2(J),\psi_1(J)]}m_1\,(\cdot,J) \;\xrightarrow[J\to\infty]{}\; 1 \;.
\end{split}\]
Therefore there exist $\bar J$ such that
\begin{equation} \label{eq: wall asymptote 5}
\big|\Delta(h,J)-\Delta\big(\!-\frac{1}{2},\,J\big)\big| \;\geq\; \frac{1}{2}\ \big|h+\frac{1}{2}\,\big| \quad\ \forall\,J\!>\bar J,\,h\!\in[\psi_2(J),\psi_1(J)] \,.
\end{equation}
Choosing $h=\gamma(J)$ in (\ref{eq: wall asymptote 5}), by (\ref{eq: wall asymptote 3}), (\ref{eq: wall asymptote 4}) one obtains that for all $J>\max\{\bar J,\,\bar J_\epsilon\}$
\[ \big|\gamma(J)+\frac{1}{2}\,\big| \,\leq\; 2\,\big|\Delta(\gamma(J),J)-\Delta\big(\!-\frac{1}{2},J\big)\big| <\, 2\epsilon \;.\qedhere\]
\endproof
\subsection{Critical exponents}
As observed in remark \ref{rk: global max m*} the global maximum point $m^*(h,J)$ is a continuous function on $(\R\times\R^+)\smallsetminus\Gamma$, but it is smooth only outside the critical point $(h_c,J_c)$.
In this section we study the behaviour of the solutions of equation (\ref{eq: ce}) near the critical point, with particular interest in the function $m^*$.
%For this purpose we consider the family of lines in the plane $(h,J)$ through the critical point $(h_c,J_c)$:
%\begin{equation} \label{eq: r_alpha}
%r_\alpha(J):=h_c+\alpha\,(J-J_c) \quad \forall\alpha\in\R  \;,
%\end{equation}
%setting also $R_\alpha:=\{(r_\alpha(J),J) \,|\, J>0\}$ and $R_\infty:=\{(h,J_c) \,|\, J>0\}$.
%Remind that by proposition \ref{prop: wall regularity} the tangent line to the wall $\overline\Gamma$ at $(h_c,J_c)$ is obtained for $\alpha=-(2m_c-1)$.\\ %To shorten the notation it will be denoted also $t$.
%To shorten the notation we denote the line tangent to the wall $\overline\Gamma$ at $(h_c,J_c)$ (see proposition \ref{prop: wall regularity}) by $t:=r_{-(2m_c-1)}$, $T:=R_{-(2m_c-1)}$.

As usual the notation $f=\mathcal{O}(g)$ as $x\to x_0$ means that there exists a neighbourhood $U$ of $x_0$ and a constant $C\in\R$ such that $|f(x)|\leq C\,|g(x)|$ for all $x\in U$. The notation $f\sim g$ as $x\to x_0$ means that $f(x)/g(x)\longrightarrow1$ as $x\to x_0$. Finally $f=o(g)$ as $x\to x_0$ means that $f(x)/g(x)\longrightarrow0$ as $x\to x_0$.

We call \textit{critical exponent} of a function $f$ at a point $x_0$ the following limit
\[ \lim_{x\to x_0}\frac{\log|f(x)-f(x_0)|}{\log|x-x_0|} \;.\]

The main result of this section is the following:
\begin{thm} \label{thm: critical exponents}
Consider the global maximum point $m^*(h,J)$ of the function $m\mapsto\vpim(m,h,J)$ defined by (\ref{eq: vp(m,h,J)}).
\begin{itemize}
\item[i)] $m^*$ is continuous on $(\R\times\R_+)\smallsetminus\Gamma$ and smooth on $(\R\times\R_+)\smallsetminus\overline\Gamma$,
where $\overline\Gamma=\Gamma\cup\{(h_c,J_c)\}$ and the ``wall'' curve $\Gamma$ is the graph of the function $\gamma$ defined by proposition \ref{prop: wall existence uniqueness}.
\item[ii)] The critical exponents of $m^*$ at the critical point $(h_c,J_c)$ are:
\[ \boldsymbol{\beta}\, = \lim_{J\to J_c+}\frac{\log|m^*(\delta(J),J)-m_c|}{\log(J-J_c)}\; =\, \frac{1}{2} \]
along any curve $h=\delta(J)$ with $\delta\in C^2([J_c,\infty[)$, $\delta(J_c)=h_c$, $\delta'(J_c)=1-2m_c$ (i.e. if the curve is tangent to the ``wall'' in the critical point);
\begin{gather*}
\frac{1}{\boldsymbol{\delta}}\, = \lim_{J\to J_c}\frac{\log|m^*(\delta(J),J)-m_c|}{\log|J-J_c|}\; =\, \frac{1}{3} \\[2pt]
\frac{1}{\boldsymbol{\delta}}\, = \lim_{h\to h_c}\frac{\log|m^*(h,\delta(h))-m_c|}{\log|h-h_c|}\; =\, \frac{1}{3}
\end{gather*}
along any curve $h=\delta(J)$ with $\delta\in C^2(\R_+)$, $\delta(J_c)=h_c$, $\delta'(J_c)\neq1-2m_c$ or along a curve $J=\delta(h)$ with $\delta\in C^2(\R)$, $\delta(h_c)=J_c$, $\delta'(h_c)=0$ (i.e. if the curve is not tangent to the ``wall'' in the critical point).
\item[iii)] Denote by $m^*(h^\pm,J):=\lim_{h'\to h\pm}m^*(h',J)$.
The critical exponent of $m^*(h^+,J)$ and $m^*(h^-,J)$ at the critical point $(h_c,J_c)$ along the ``wall'' $h=\gamma(J)$ is still
\begin{gather*}
\boldsymbol{\beta}\, = \lim_{J\to J_c+}\frac{\log|m^*(\gamma(J)^+,J)-m_c|}{\log(J-J_c)}\; =\, \frac{1}{2} \\[2pt]
\boldsymbol{\beta}\, = \lim_{J\to J_c+}\frac{\log|m^*(\gamma(J)^-,J)-m_c|}{\log(J-J_c)}\; =\, \frac{1}{2}
\end{gather*}
\end{itemize}
\end{thm}

\proof As observed in remark \ref{rk: global max m*}, the global maximum point $m^*$ is expressed piecewise using the two local maximum points $\mu_1$, $\mu_2$ and inherits their continuity property outside $\Gamma$ and their regularity properties outside $\overline\Gamma$.
Thus part \textit{i)} of the theorem is already proved by proposition \ref{prop: continuity of stationary points of vp}.\\
The proof of the other parts of the theorem, regarding the behaviour of $m^*$ at the critical point $(h_c,J_c)$, is given in several steps. We start with the following lemma which will be useful in the whole subsection to bound the behaviour of the solutions of equation (\ref{eq: ce}).

\begin{lem} \label{lem: flex exponent}
Consider the inflection points $\phi_1,\,\phi_2$ of $\vpim$ defined by (\ref{eq: phi(h,J)}).
Their behaviour at the critical point $(h_c,J_c)$ along any curve $\delta\in C^1([J_c,\infty[)$, with $\delta(J_c)=h_c$, is
\[ \frac{\phi_1(\delta(J),J)-m_c}{\sqrt{J-Jc}}\;\xrightarrow[J\to J_c+]{}\,-C \;,\quad
\frac{\phi_2(\delta(J),J)-m_c}{\sqrt{J-Jc}}\;\xrightarrow[J\to J_c+]{}\;C \;.\]
where $C=\sqrt[4]{2}/(2J_c)>0$.
%For $J>J_c$ consider the local maximum points $m_1,\,m_2$ of $\vpim$ defined by proposition \ref{prop: stationary points of vp}. As a consequence for $i=1,2$
%\[ (J-J_c)^{\frac{1}{2}} = \mathcal{O}\big(\,m_i(r_\alpha(J),J)-m_c\,\big) \quad\text{as }J\to J_c+ \;.\]
\end{lem}

\proof
For $i=1,2$ and $J\geq J_c$ definition (\ref{eq: phi(h,J)}), observing that %by (\ref{eq: xic})
$(2m_c-1)J = -h_c+(2m_c-1)\,(J-J_c)+\xi_c\,$, gives
\[ 2J\,\big(\phi_i(\delta(J),J)-m_c\big) \,=\, \frac{1}{2}\log a_i(J)-\xi_c - (\delta(J)-h_c)-(2m_c-1)(J-J_c) \;.\]
%\,&=\, J-\delta(J)+\frac{1}{2}\log a_i(J) -2J\,m_c \\
Now the definition (\ref{eq: a1,a2(J)}) may be rewritten as
\[ a_i(J) \,=\, \underbrace{(2J-2-\frac{1}{8J})}_{=:\,b(J)} \;\mp\; \underbrace{4\,(\frac{1}{2}-\frac{1}{8J})\,\sqrt{J-\tfrac{3-2\sqrt{2}}{4}}}_{=:\,c(J)}\ \sqrt{J-Jc} \;.\]
Thus, exploiting $\log(x+y)=\log x+\log(1+y/x)=\log x+y/x+\mathcal{O}((y/x)^2)$ as $y/x\to0$, $\frac{1}{2}\,\log b(J_c)=\xi_c$ and $\log b(J)$ differentiable at $J=J_c$,
\[\begin{split}
\frac{1}{2}\log a_i(J) - \xi_c \,&=\,
\frac{1}{2}\,\frac{\log b(J)-\log b(J_c)}{(J-J_c)}\;(J-J_c) \,\mp\, \frac{1}{2}\,\frac{c(J)}{b(J)}\;\sqrt{J-Jc} \,+ \mathcal{O}(J-J_c) \\
&=\, \mp\,\frac{1}{2}\,\frac{c(J)}{b(J)}\;\sqrt{J-J_c} \,+ \mathcal{O}(J-J_c) \;.
\end{split}\]
%\frac{1}{2}\log b(J)-\xi_c + \frac{1}{2}\log\big(\,1\pm\frac{c(J)}{b(J)}\;\sqrt{J-Jc}\,\big) \\
To conclude put things together and use also $\delta$ differentiable at $J_c$: %and $\frac{1}{2}\,\frac{c(J)}{b(J)}\longrightarrow\sqrt[4]{2}$ as $J\to J_c$:
\[\begin{split}
2J\,\frac{\phi_i(\delta(J),J)-m_c}{\sqrt{J-Jc}} \,&=\,
\frac{\frac{1}{2}\,\log a_i(J)-\xi_c}{\sqrt{J-J_c}}\, - \frac{\delta(J)-h_c}{\sqrt{J-J_c}}\, -(2m_c-1)\sqrt{J-J_c} \\[2pt]
&=\, \mp\frac{1}{2}\,\frac{c(J)}{b(J)} + \mathcal{O}(\sqrt{J-J_c}\;)
\,\xrightarrow[J\to J_c+]{}\; \pm\sqrt[4]{2} \;. \qedhere
\end{split}\]
\endproof

In the following proposition we find the fundamental equation characterizing the behaviour of the solutions of equation (\ref{eq: ce}) near the critical point $(h_c,J_c)$.

\begin{prop} \label{prop: critical equation}
Here for $h\in\R,\,J>0$ let $m=m(h,J)$ be any solution of the consistency equation (\ref{eq: ce}):
\[ m=g\big((2m-1)J+h\big) \;.\]
Then $m$ is continuous at $(h_c,J_c)$ and furthermore, setting $\xi:=(2m-1)J+h$, it satisfies
\begin{equation} \label{eq: critical equation}
(\xi-\xi_c)^3 - \kappa_1\,(J-J_c)\,(\xi-\xi_c) - \kappa_2\,\rho(h,J) + \mathcal{O}\big((\xi-\xi_c)^4\big) = 0
\end{equation}
as $(h,J)\to(h_c,J_c)$, where we set $\kappa_1:=3\,\frac{J_c}{J}\,(2-m_c)$, $\kappa_2:=3\,\frac{J_c^2}{J}\,(2-m_c)$ and
\begin{equation} \label{eq: rho}
\rho(h,J):=h-h_c+(2m_c-1)(J-J_c) \;.
\end{equation}
\end{prop}

\proof
\textit{I.} First show that $m$ is continuous at $(h_c,J_c)$. Exploit equation (\ref{eq: ce}) for $m(h,J)$ and use continuity and monotonicity of $g$: as $(h,J)\to(h_c,J_c)$
\begin{gather*}
\limsup m(h,J) = \limsup g\big((2\,m(h,J)-1)\,J+h\big) = g\big((2\limsup m(h,J)-1)\,J_c+h_c\big) \;,\\[2pt]
\liminf m(h,J) = \liminf g\big((2\,m(h,J)-1)\,J+h\big) = g\big((2\liminf m(h,J)-1)\,J_c+h_c\big) \;.
\end{gather*}
Thus $\limsup m(h,J)$ and $\liminf m(h,J)$ are both solution of equation $\mu=g\big((2\mu+1)J_c+h_c\big)$. But this solution is unique by proposition \ref{prop: stationary points of vp}, and it is $m_c$ by remark \ref{rk: mc=m(hc,Jc)}. Therefore
\[ \limsup_{(h,J)\to(h_c,J_c)} m(h,J) \,= \liminf_{(h,J)\to(h_c,J_c)} m(h,J) \,=\, m_c \;.\]
\textit{II.} Make a Taylor expansion of the smooth function $g$ at the point $\xi_c$ (see (\ref{eq: g(h)}), (\ref{eq: xic})). By identities (\ref{eq: g' function of g}), (\ref{eq: g'' function of g}), (\ref{eq: g''' function of g}) and since $g(\xi_c)=m_c$ it is easy to find
\begin{equation} \label{eq: taylor}
g(\xi) \,=\, m_c + \frac{1}{2J_c}\,(\xi-\xi_c) - \frac{1}{6J_c^2(2-m_c)}\,(\xi-\xi_c)^3 + \mathcal{O}\big((\xi-\xi_c)^4\big)
\end{equation}
as $\xi\to\xi_c$.
Now choose $\xi:=(2m-1)J+h$. Then $g(\xi)=m$ and
\begin{equation} \label{eq: xi - m}
\xi-\xi_c \,=\, \rho(h,J) + 2J\,(m-m_c) \;,
\end{equation}
where $\rho(h,J):=h-h_c+(2m_c-1)(J-J_c)$.
Now (\ref{eq: critical equation}) follows from (\ref{eq: taylor}).
%Thus identity (\ref{eq: taylor}) rewrites as
%\[ (\xi-\xi_c)^3 - \kappa_1\,(J-J_c)\,(\xi-\xi_c) - \kappa_2\,\rho(h,J) + \mathcal{O}\big((\xi-\xi_c)^4\big) = 0 \]
%where $\kappa_1:=3\,\frac{J_c}{J}\,(2-m_c)$, $\kappa_2:=3\,\frac{J_c^2}{J}\,(2-m_c)$.
\endproof

Hereafter we will exploit equation (\ref{eq: critical equation}) and lemma \ref{lem: flex exponent} in order to obtain results on the behaviour near the critical point.
Next corollary gives a first bound for the critical exponents.

\begin{cor} \label{cor: bounds for critical exponents}
Here for $h\in\R,\,J>J_c$ let $m=m(h,J)$ be any solution of the consistency equation (\ref{eq: ce}).\\[2pt]
\textit{1)} There exist $r_1>0,\,C_1<\infty$ such that for all $(h,J)\in B\big((h_c,J_c),r_1\big)$ with $J>J_c$
\[ |m-m_c| \,\leq\, C_1\big(\, |h-h_c|^{\frac{1}{3}} + |J-J_c|^{\frac{1}{3}} \big) \;.\]
\textit{2)} Assume that $m$ pointwise coincides with one of the local maximum points $m_1$, $m_2$ (see proposition \ref{prop: stationary points of vp}). There exist $r_2>0,\,C_2>0$ such that for all $(h,J)\in B\big((h_c,J_c),r_2\big)$ with $J>J_c$ and $h=\delta(J)$ for some $\delta\in C^1([J_c,\infty[),\,\delta(J_c)=h_c$
\[ |m-m_c| \,\geq\, C_2\,|J-J_c|^{\frac{1}{2}} \;.\]
%The constants $C_1,\,C_2$ do not depend on anything, not even the choice of the curve $\delta$.
\end{cor}

\proof
\textit{1)} Set $\xi:=(2m-1)J+h$. By proposition \ref{prop: critical equation}, $\xi$ satisfies equation (\ref{eq: critical equation}), which can be treated as a third degree algebraic equation in $\xi-\xi_c$:
\[ (\xi-\xi_c)^3 \underbrace{- \kappa_1\,(J-J_c)}_{=:\,p}\,(\xi-\xi_c) \underbrace{- \kappa_2\,\rho(h,J) + \mathcal{O}\big((\xi-\xi_c)^4\big)}_{=:\,q} = 0 \;.\]
Analyse the real solutions of this equation. %\ref{propA: third degree equations}.
Set $\Delta:=(\frac{q}{2})^2+(\frac{p}{3})^3$ and observe that $(\frac{q}{2})^2>0$ while $(\frac{p}{3})^3<0$ as we are assuming $J>J_c$.\\[2pt]
\textit{i.} If $\Delta>0$, the only real solution of (\ref{eq: critical equation}) is
\[ \xi-\xi_c = u_++u_- \quad\text{with }\,u_\pm=\sqrt[3]{-\frac{q}{2}\pm\sqrt[2]{\Delta}} \;.\]
On the other hand
\[ \Delta>0 \ \Rightarrow\ \big(\frac{p}{3}\,\big)^3=\mathcal{O}\big(\big(\frac{q}{2}\,\big)^2\big)\ \Rightarrow\  \Delta=\mathcal{O}\big(\big(\frac{q}{2}\,\big)^2\big) \;.\]
Therefore, reminding also definition (\ref{eq: rho}),
\[ \xi-\xi_c \,=\, \mathcal{O}\big(\big(\frac{q}{2}\,\big)^\frac{1}{3}\big) \,=\,
\mathcal{O}\big((h-h_c)^\frac{1}{3}\big) + \mathcal{O}\big((J-J_c)^\frac{1}{3}\big) + \mathcal{O}\big((\xi-\xi_c)^\frac{4}{3}\big)\;,\]
hence $\xi-\xi_c=\mathcal{O}\big((h-h_c)^\frac{1}{3}\big) + \mathcal{O}\big((J-J_c)^\frac{1}{3}\big)$ because $(\xi-\xi_c)^{\frac{4}{3}-1}\rightarrow0$ as $(h,J)\to(h_c,J_c)\,$.\\[2pt]
\textit{ii.} If $\Delta=0$ or $\Delta<0$ there are respectively two or three distinct real solutions of (\ref{eq: critical equation}) and, from their explicit form, it is immediate to see  that they all satisfy
\[ \xi-\xi_c \,=\, \mathcal{O}\big(\sqrt[2]{-\frac{p}{3}}\;\big) \,=\, \mathcal{O}\big((J-J_c)^\frac{1}{2}\big) \;.\]
%\textit{ii.} If $\Delta<0$, there are three real solutions of (\ref{eq: critical equation}):
%\[ \xi-\xi_c = 2\,\sqrt[2]{-\frac{p}{3}}\ \cos\frac{\theta+2\pi k}{3} \quad\text{with }\theta=\arg\big(\!-\frac{q}{2}+i\sqrt{-\Delta}\,\big)\,,\ k\in\{0,1,2\}\;.\]
%Therefore in any case
%\[ \xi-\xi_c \,=\, \mathcal{O}\big(\sqrt[2]{-\frac{p}{3}}\;\big) \,=\, \mathcal{O}\big((J-J_c)^\frac{1}{2}\big) \;.\]
%\textit{iii.} If $\Delta=0$, there are two real solutions of (\ref{eq: critical equation}):
%\[ \xi-\xi_c = \epsilon\,\sqrt[3]{\frac{q}{2}} \quad\text{with }{\epsilon\in\{-2,1\}} \;.\]
%On the other hand
%\[ \Delta=0 \ \Rightarrow\ \big(\frac{p}{3}\,\big)^3=-\big(\frac{q}{2}\,\big)^2\ \Rightarrow\  \big|\frac{q}{2}\big|^\frac{1}{3}=\big(\!-\frac{p}{3}\big)^\frac{1}{2} \;.\]
%Therefore in any case
%\[ \xi-\xi_c \,=\, \mathcal{O}\big(\big(-\frac{p}{3}\,\big)^\frac{1}{2}\big) \,=\, \mathcal{O}\big((J-J_c)^\frac{1}{2}\big) \;.\]
Conclude that for any possible value of $\Delta$,
\[ \xi-\xi_c=\mathcal{O}\big((h-h_c)^\frac{1}{3}\big) + \mathcal{O}\big((J-J_c)^\frac{1}{3}\big) \;.\]
Now, as observed in (\ref{eq: xi - m}), $\xi-\xi_c=h-h_c+(2m_c-1)(J-J_c)+2J\,(m-m_c)$. Therefore also
$ m-m_c=\mathcal{O}\big((h-h_c)^\frac{1}{3}\big) + \mathcal{O}\big((J-J_c)^\frac{1}{3}\big)\,$,
and this concludes the proof of the first statement.\\[2pt]
\textit{2)} Now consider the two maximum points $m_1,\,m_2$. By proposition \ref{prop: stationary points of vp}
\[ m_1 < \phi_1 < \phi_2 < m_2 \]
where $\phi_1,\,\phi_2$ are the inflection points defined by (\ref{eq: phi(h,J)}). Hence applying lemma \ref{lem: flex exponent} one finds:
\begin{gather*}
\frac{m_2-m_c}{\sqrt{J-J_c}} \,>\, \frac{\phi_2-m_c}{\sqrt{J-J_c}} \,\longrightarrow\; C \,, \quad
\frac{m_c-m_1}{\sqrt{J-J_c}} \,>\, \frac{m_c-\phi_1}{\sqrt{J-J_c}} \,\longrightarrow\; C \,,
\end{gather*}
as $J\to J_c+$ and $h=\delta(J)$ with $\delta(J_c)=h_c$ and $\delta$ differentiable in $J_c$.
And this proves the second statement.
\endproof

The next lemma tells us in which region of the plane $(h,J)$ described by figure \ref{fig: stationary points regions} a curve passing through the point $(h_c,J_c)$ lies.

\begin{lem} \label{lem: lines and psi}
Let $\delta\in C^2([J_c,\infty[)$ such that $\delta(J_c)=h_c$, $\delta'(J_c)=:\alpha$.
There exists $r>0$ such that for all $J\in\,]J_c,J_c+r[$
\begin{itemize}
\item if $\alpha=1-2m_c$, $\;\psi_2(J)<\delta(J)<\psi_1(J)\,$;
\item if $\alpha<1-2m_c$, $\;\delta(J)<\psi_2(J)\,$;
\item if $\alpha>1-2m_c$, $\;\delta(J)>\psi_1(J)\,$.
\end{itemize}
\end{lem}

\proof
\textit{I.} %By definition (\ref{eq: psi(J)}) for $i=1,2$ and $J\geq J_c$
%\[ \psi_i(J) \,=\, J+\frac{1}{2}\log a_i(J) -2J\,g(\frac{1}{2}\log a_i(J)) \;.\]
Observe that $a_i(J)$ is continuous for $J\geq J_c$ and smooth for $J>J_c$. Moreover $g'(\frac{1}{2}\log a_i(J))=\frac{1}{2J}$ by definition (\ref{eq: a1,a2(J)}) and lemma \ref{lemA: g'<c}, and $g(\frac{1}{2}\log a_i(J_c))=g(\xi_c)=m_c$ by definition (\ref{eq: xic}) and remark \ref{rk: mc=m(hc,Jc)}.
Then differentiating definition (\ref{eq: psi(J)}) at $J>J_c$,
\[ \psi_i'(J) \,=\, 1-2\,g(\frac{1}{2}\log a_i(J)) + \frac{1}{2}\,\frac{a_i'(J)}{a_i(J)}\,\big(\underbrace{1-2J\,g'(\frac{1}{2}\log a_i(J))}_{=\,0}\big) \,\xrightarrow[J\to J_c]{}\, 1-2m_c \;. \\[-5pt]\]
%\[ \psi_i'(J) = 1-2\,g(\frac{1}{2}\log a_i(J)) + \frac{1}{2}\frac{a_i'(J)}{a_i(J)}\big(1-2J\,g'(\frac{1}{2}\log a_i(J))\big) = 1-2\,g(\frac{1}{2}\log a_i(J))\]
%Now using $\frac{1}{2}\log a_i(J_c)=\xi_c$ and $g(\xi_c)=m_c$ (see remark \ref{rk: mc=m(hc,Jc)}),
%\[ 1-2\,g(\frac{1}{2}\log a_i(J)) \,\xrightarrow[J\to J_c+]{}\, 1-2\,g(\xi_c) \,=\, 1-2m_c \;.\]
Hence an immediate application of the mean value theorem %(lemma \ref{lemA: elementary property of derivative})
shows that for $i=1,2$ there exits $\psi_i'(J_c)=1-2m_c\,$.\\[2pt]
\textit{II.} Differentiating definition (\ref{eq: a1,a2(J)}) at $J>J_c$ shows that $a_1'(J)\rightarrow-\infty$, $a_2'(J)\rightarrow+\infty$ as $J\to J_c+$, while $a_i(J)\rightarrow 2\sqrt{2}-2$ as $J\to J_c$. Hence
\[ \psi_i''(J) \,=\, -\,g'(\frac{1}{2}\log a_i(J))\,\frac{a_i'(J)}{a_i(J)}  \,=\, -\frac{1}{2J}\,\frac{a_i'(J)}{a_i(J)} \;\xrightarrow[J\to J_c+]{}\;
\begin{cases}
+\infty & \text{for }i=1 \\
-\infty & \text{for }i=2
\end{cases} \;.\]
%Hence an immediate application of the mean value theorem %(lemma \ref{lemA: elementary property of derivative})
%shows that $\psi_1''(J_c)=+\infty\,$, $\psi_2''(J_c)=-\infty\,$.\\[2pt]
The result is provided comparing the first order Taylor expansions at $J_c$ with Lagrange remainder of $\psi_1$, $\psi_2$ and $\delta$.
%\textit{III.} By \textit{I.}, \textit{II.} and by (\ref{eq: hc}) the Taylor expansion of $\psi_1$, $\psi_2$ at $J_c$ with remainder in the Lagrange form is
%\[ \psi_i(J) \,=\, h_c - (2m_c-1)\,(J-J_c) + \psi_i''(\bar J_i)\,(J-J_c)^2 \quad\text{with }\bar J_i\in\,]J_c,J[ \]
%and $\psi_1''(\bar J_1)\longrightarrow+\infty$, $\psi_2''(\bar J_2)\longrightarrow+\infty$ as $J\to J_c$.
%On the other hand the Taylor expansion of $\delta$ at $J_c$ is
%\[ \delta(J) \,=\, h_c + \alpha\,(J-J_c) + \delta''(\bar J)\,(J-J_c)^2 \quad\text{with }\bar J\in\,]J_c,J[ \]
%and $\delta''(\bar J)\longrightarrow\delta''(J_c)\in\R$ as $J\to J_c$.
%The result is provided comparing $\psi_1$, $\psi_2$ with $\delta$ in this form, for $J$ sufficiently small.
\endproof

The next proposition describes the behaviour near $(h_c,J_c)$ of the two local maximum points $\mu_1,\,\mu_2$ defined in proposition \ref{prop: continuity of stationary points of vp}.
The proof of part \textit{ii)} of the theorem \ref{thm: critical exponents} follows immediately.

\begin{prop} \label{prop: critical exponents}
Let $(h,J)\to(h_c,J_c)$ along a curve $h=\delta(J)$ with $\delta\in C^2(\R_+)$, $\delta(J_c)=h_c$, $\delta'(J_c)=:\alpha$ or along a curve $J=\delta(h)$ with $\delta\in C^2(\R)$, $\delta(h_c)=J_c$, $\delta'(h_c)=0$, then
\begin{gather*} \mu_1(h,J)-m_c \,\sim\, \begin{cases}
\,-\,C\,(J-J_c)^{\frac{1}{2}} &  \text{if }h=\delta(J),\,\alpha=1-2m_c\text{ and }J>J_c \\[2pt]
\,C_\alpha\,(J-J_c)^{\frac{1}{3}} & \text{if }h=\delta(J),\,\alpha<1-2m_c\\[2pt]
\,C_\infty\,(h-h_c)^{\frac{1}{3}} &  \text{if }J=\delta(h)
\end{cases} \\[4pt]
\mu_2(h,J)-m_c \,\sim\, \begin{cases}
\,C\,(J-J_c)^{\frac{1}{2}} &  \text{if }h=\delta(J),\,\alpha=1-2m_c\text{ and }J>J_c \\[2pt]
\,C_\alpha\,(J-J_c)^{\frac{1}{3}} & \text{if }h=\delta(J),\,\alpha>1-2m_c\\[2pt]
\,C_\infty\,(h-h_c)^{\frac{1}{3}} &  \text{if }J=\delta(h)
\end{cases} \end{gather*}
where $C=\frac{1}{2J_c}\,\sqrt{3(2-m_c)}\,$, $C_\alpha=\frac{1}{2J_c}\,\sqrt[3]{\frac{3}{2}J_c(2-m_c)(2m_c-1+\alpha)}\,$,  $C_\infty=\frac{1}{2J_c}\,\sqrt[3]{3J_c(2-m_c)}\,$.
To complete the cases, along the line $h=h_c+(1-2m_c)(J-J_c)$, when $J\leq J_c$
\[ \mu_1(h,J)=\mu_2(h,J)=m_c .\]
\end{prop}

\proof
Fix $(h,J)$ on the curve given by the graph of $\delta$ and in the rest of the proof denote by $m$ a solution of the consistency equation (\ref{eq: ce}), i.e. $m=g\big((2m-1)J+h\big)$. Furthermore when necessary $m$ is assumed to be a local maximum point of $\vpim$.
Set $\xi:=(2m-1)J+h$. By proposition \ref{prop: critical equation}, $\xi-\xi_c\rightarrow0$ as $(h,J)\to(h_c,J_c)$ and it satisfies (\ref{eq: critical equation}). Solve this equation in the different cases.\\[2pt]
\textit{i)} Suppose $h=\delta(J)$ with $\alpha=1-2m_c$. Hence $h-h_c=(1-2m_c)(J-J_c)+\mathcal{O}\big((J-J_c)^2\big)$. Observe that by (\ref{eq: rho}), (\ref{eq: xi - m})
\[ \rho(h,J)=\mathcal{O}\big((J-J_c)^2\big) \quad\text{and}\quad \xi-\xi_c=2J\,(m-m_c)+\mathcal{O}\big((J-J_c)^2\big) \;.\]
Hence equation (\ref{eq: critical equation}) becomes
\[ (\xi-\xi_c)^3 - \kappa_1\,(J-J_c)\,(\xi-\xi_c) + \mathcal{O}\big((J-J_c)^2\big) + \mathcal{O}\big((\xi-\xi_c)^4\big) = 0 \;.\]
Observe that if $J>J_c$ by corollary \ref{cor: bounds for critical exponents} part \textit{2)},
\[ (J-J_c)^\frac{1}{2} \,=\, \mathcal{O}(\xi-\xi_c) \;;\]
%\,=\, \mathcal{O}(m-m_c)
therefore when $J>J_c$ the previous equation rewrites
\[ (\xi-\xi_c)^3 - \kappa_1\,(J-J_c)\,(\xi-\xi_c) + \mathcal{O}\big((\xi-\xi_c)^4\big) = 0 \;.\]
This one simplifies in
\[ \xi=\xi_c \quad\text{or}\quad (\xi-\xi_c)^2 - \kappa_1\,(J-J_c) + \mathcal{O}\big((\xi-\xi_c)^3\big) = 0 \;,\]
giving $\xi=\xi_c$ or, as we are assuming $J>J_c$,
\[ \xi-\xi_c \,=\, \pm\, \sqrt{\kappa_1}\;(J-J_c)^{\frac{1}{2}} + \mathcal{O}\big((\xi-\xi_c)^{\frac{3}{2}}\big) \;.\]
This entails
\[ m-m_c \,=\, \pm\, \frac{\sqrt{\kappa_1}}{2J}\;(J-J_c)^{\frac{1}{2}} + \mathcal{O}\big((J-J_c)^2\big) + \mathcal{O}\big((m-m_c)^{\frac{3}{2}}\big) \]
and dividing both sides by $m-m_c$, since $(m-m_c)^\frac{1}{2}\rightarrow0$, one finds
\begin{equation} \label{eq: critical for alpha=-(2mc-1)}
m-m_c \,\sim\, \pm\,\frac{\sqrt{\kappa_1}}{2J}\,(J-J_c)^{\frac{1}{2}} \;.\\[2pt]
\end{equation}
\textit{ii)} Suppose $J=\delta(h)$ with $\delta'(h_c)=0$. Hence $J-J_c=\mathcal{O}\big((h-h_c)^2\big)$. (\ref{eq: rho}) and (\ref{eq: xi - m}) give
\[ \rho(h,J)=h-h_c+\mathcal{O}\big((h-h_c)^2\big) \quad\text{and}\quad \xi-\xi_c=2J\,(m-m_c)+h-h_c+\mathcal{O}\big((h-h_c)^2\big) \,.\]
Hence equation (\ref{eq: critical equation}) becomes
\[ (\xi-\xi_c)^3 - \kappa_2\,(h-h_c) + \mathcal{O}\big((h-h_c)^2\big) + \mathcal{O}\big((\xi-\xi_c)^4\big) = 0 \;.\]
giving
\[ \xi-\xi_c \,=\, \sqrt[3]{\kappa_2}\;(h-h_c)^{\frac{1}{3}} + \mathcal{O}\big((h-h_c)^\frac{2}{3}\big) + \mathcal{O}\big((\xi-\xi_c)^{\frac{4}{3}}\big) \;.\]
This entails
\[ m-m_c \,=\, \frac{\sqrt[3]{\kappa_2}}{2J}\;(h-h_c)^{\frac{1}{3}} + \mathcal{O}\big((h-h_c)^\frac{2}{3}\big) + \mathcal{O}\big((m-m_c)^\frac{4}{3}\big) \]
and dividing both sides by $m-m_c$, since $(m-m_c)^\frac{1}{3}\rightarrow0$, one finds
\begin{equation} \label{eq: critical for alpha=infty}
m-m_c \,\sim\, \frac{\sqrt[3]{\kappa_2}}{2J}\;(h-h_c)^{\frac{1}{3}} \;.\\[2pt]
\end{equation}
\textit{iii)} Suppose $h=\delta(J)$ with $\alpha\neq1-2m_c$. Hence $h-h_c=\alpha\,(J-J_c)+\mathcal{O}\big((J-J_c)^2\big)$. Observe that by (\ref{eq: rho}), (\ref{eq: xi - m})
\begin{gather*}
\rho(h,J)=(\alpha+2m_c-1)(J-J_c)+\mathcal{O}\big((J-J_c)^2\big) \;,\\[2pt]
\xi-\xi_c = 2J\,(m-m_c) + (\alpha+2m_c-1)(J-J_c) + \mathcal{O}\big((J-J_c)^2\big) \,.
\end{gather*}
Hence equation (\ref{eq: critical equation}) becomes
\[ (\xi-\xi_c)^3 \underbrace{- \kappa_1\,(J-J_c)}_{=:\,p}\;(\xi-\xi_c) \underbrace{- \kappa_2\,(\alpha+2m_c-1)\,(J-J_c) + \mathcal{O}\big((J-J_c)^2\big) + \mathcal{O}\big((\xi-\xi_c)^4\big)}_{=:\,q} = 0 \;. \]
This third order equation has $\Delta:=(\frac{q}{2})^2+(\frac{p}{3})^3>0$ for $|J-J_c|$ small enough, indeed if $J<J_c$ then $p>0$, while if $J>J_c$ then by corollary \ref{cor: bounds for critical exponents} part \textit{1)} $(\xi-\xi_c)^4=\mathcal{O}\big((J-J_c)^\frac{4}{3}\big)=o(J-J_c)$ hence
\[\begin{split}
&q = -\kappa_2\,(\alpha+2m_c-1)\,(J-J_c) + o\big(J-J_c\big) \quad\Rightarrow\\
&\big(\frac{q}{2}\,\big)^2+\big(\frac{p}{3}\,\big)^3 \,=\,
\frac{\kappa_2^2}{4}\,(\underbrace{\alpha+2m_c-1}_{\neq\,0})^2\,(J-J_c)^2\,(1+o(1)) - \frac{\kappa_1^3}{27}\,(J-J_c)^3 \,>\, 0 \;.
\end{split}\]
Then, using Cardano's formula for cubic equations: %(see proposition \ref{propA: third degree equations}):
$ \xi-\xi_c = u_+ + u_- $ with
%with (using $\sqrt[n]{a+b} \leq \sqrt[n]{a}+\sqrt[n]{b}$ for a certain $0\leq B\leq\sqrt[n]{|b|}$, when $a>0$ and $a+b>0$)
\[ u_\pm \,=\, \sqrt[3]{-\frac{q}{2}\pm\sqrt[2]{\big(\frac{q}{2}\,\big)^2+\big(\frac{p}{3}\,\big)^3}}\ =\,
 \sqrt[3]{-\frac{q}{2}\pm\big|\frac{q}{2}|}\; +\mathcal{O}\big(\big|\frac{p}{3}\big|^\frac{1}{2}\big) \;; \]
%\sqrt[3]{-\frac{q}{2}\pm\big|\frac{q}{2}| +\mathcal{O}\big(\big|\frac{p}{3}\big|^\frac{3}{2}\big)} \,=\,
%
%\[\begin{split}
%u_\pm \,&=\, \sqrt[3]{-\frac{q}{2}\pm\sqrt[2]{\big(\frac{q}{2}\,\big)^2+\big(\frac{p}{3}\,\big)^3}}\ =\,
%\sqrt[3]{-\frac{q}{2}\pm\big|\frac{q}{2}\big|\,\sqrt[2]{1+\frac{4\,p^3}{27\,q^2}}}\ =\\
%&=\, \sqrt[3]{-\frac{q}{2}}\ \;\sqrt[3]{1\mp\sgn\big(\frac{q}{2}\big)\,\sqrt[2]{1+\frac{4\,p^3}{27\,q^2}}} \;;
%\end{split}\]
hence
\[\begin{split}
\xi-\xi_c \,&=\, \sqrt[3]{-q}\, + \mathcal{O}\big(\big|\frac{p}{3}\big|^\frac{1}{2}\big) \,=\\
&=\, \sqrt[3]{\kappa_2\,(\alpha+2m_c-1)}\;(J-J_c)^\frac{1}{3} + \mathcal{O}\big((J-J_c)^\frac{2}{3}\big) + \mathcal{O}\big((\xi-\xi_c)^\frac{4}{3}\big) + \mathcal{O}\big((J-J_c)^\frac{1}{2}\big) \;.
\end{split}\]
This entails
\[ m-m_c \,=\,  \frac{\sqrt[3]{\kappa_2\,(\alpha+2m_c-1)}}{2J}\;(J-J_c)^\frac{1}{3} + \mathcal{O}\big((J-J_c)^\frac{1}{2}\big) + \mathcal{O}\big((m-m_c)^\frac{4}{3}\big) \]
and dividing both sides by $m-m_c$, since $(m-m_c)^\frac{1}{2}\rightarrow0$, one finds
\begin{equation} \label{eq: critical for other alpha}
m-m_c \,\sim\,  \frac{\sqrt[3]{\kappa_2\,(\alpha+2m_c-1)}}{2J}\;(J-J_c)^\frac{1}{3} \;.\\[2pt]
\end{equation}
Now by propositions \ref{prop: stationary points of vp}, \ref{prop: continuity of stationary points of vp} and lemma \ref{lem: lines and psi}, $\mu_1$ and $\mu_2$ are solutions of the consistency equation (\ref{eq: ce}) defined near $(h_c,J_c)$ along the curves $h=\delta(J)$ respectively with $\alpha\leq1-2m_c$ and $\alpha\geq1-2m_c$.
Moreover for $\alpha=1-2m_c$ and $J>J_c$ sufficiently small, by lemma \ref{lem: flex exponent},
\[ \mu_2-m_c>\phi_2-m_c>0 \quad\text{while}\quad \mu_1-m_c<\phi_1-m_c<0 \;.\]
These facts together with (\ref{eq: critical for alpha=-(2mc-1)}), (\ref{eq: critical for alpha=infty}), (\ref{eq: critical for other alpha}) allow to conclude the proof.
\endproof

The previous proposition describes the critical behaviour of the local maximum points along curves of class $C^2$.
Notice that ``the wall'' $\overline\gamma$ belongs to $C^1([J_c,\infty[)\,\cap\,C^\infty(]J_c,\infty[)$ by proposition \ref{prop: wall regularity}, but we did not manage to prove that it is $C^2$ up to $J_c$.
Anyway we are interested in the behaviour along this curve of discontinuity, which separates two different states of the system, therefore we will study it in the following proposition.

\begin{prop} \label{prop: critical exponents along wall}
Consider the ``wall'' curve $h=\overline\gamma(J)$ defined by (\ref{eq: gamma}) and proposition \ref{prop: wall existence uniqueness}.
There exist $r>0$, $C_1<\infty$, $C_2>0$ such that for all $J\in\,]J_c,J_c+r[\,$.
\[ C_2 \leq \frac{\mu_2(\overline\gamma(J),J)-m_c}{\sqrt{J-J_c}} \leq C_1\,, \quad C_2 \leq \frac{m_c-\mu_1(\overline\gamma(J),J)}{\sqrt{J-J_c}} \leq C_1 \]
\end{prop}

\proof
Observe that by definition, on the curve $h=\overline\gamma(J),\,J\geq J_c$, both the local maximum points $\mu_1(h,J),\,\mu_2(h,J)$ exist.

As $\overline\gamma\in C^1([J_c,\infty[)$ (see proposition \ref{prop: wall regularity}), the existence of the lower bound $C_2>0$ is guaranteed by corollary \ref{cor: bounds for critical exponents} part \textit{2)}.

Only the existence of an upper bound $C_1<\infty$ has to be proven. Fix $J>J_c$ and shorten the notation by $m_i=m_i(\overline\gamma(J),J)=\mu_i(\overline\gamma(J),J)$ and $\xi_i:=(2m_i-1)\,J+\gamma(J)$ for $i=1,2$.
By proposition \ref{prop: critical equation}, $\xi_1,\,\xi_2$ satisfy equation (\ref{eq: critical equation}).
The Taylor expansion with Lagrange remainder of $\overline\gamma$ is (see proposition \ref{prop: wall regularity})
\[ \gamma(J) = h_c + (1-2m_c)\,(J-J_c) + \gamma''(\bar J)\,(J-J_c)^2 \,,\quad\text{with }\bar J\in\,]J_c,J[ \;;\]
notice $\gamma''(\bar J)\,(J-J_c)^2$ is not necessarily a $\mathcal{O}\big((J-J_c)^2\big)$, because we do not know the behaviour of $\gamma''$ as $J\to J_c$, but for sure it is a $o(J-J_c)$ as $J\to J_c$.\\
Thus (see identities (\ref{eq: rho}), (\ref{eq: xi - m})):
\[ \rho(h,J) = \gamma''(\bar J)\,(J-J_c)^2 \quad\text{and}\quad \xi_i-\xi_c = 2J\,(m_i-m_c)+\gamma''(\bar J)\,(J-J_c)^2 \]
and equation (\ref{eq: critical equation}) becomes:
\[ (\xi_i-\xi_c)^3 - \kappa_1\,(J-J_c)\,(\xi_i-\xi_c) - \kappa_2\,\gamma''(\bar J)\,(J-J_c)^2 + \mathcal{O}\big((\xi_i-\xi_c)^4\big) = 0 \;,\]
which entails
\begin{multline} \label{eq: critical equation for gamma}
(m_i-m_c)^3 - \frac{\kappa_1}{(2J)^2}\,(J-J_c)\,(m_i-m_c) - \frac{\kappa_2}{(2J)^3}\,\gamma''(\bar J)\,(J-J_c)^2\,(1+o(1)) +\\
+ \mathcal{O}\big((m_i-m_c)^4\big) = 0 \,.
\end{multline}
Distinguish two cases.\\[2pt]
\textit{1)} If $\gamma''(\bar J)\,(J-J_c)^2 = \mathcal{O}\big((m_i-m_c)^4\big)$ (along a sequence), then (\ref{eq: critical equation for gamma}) rewrites
\begin{equation} \label{eq: critical equation for gamma 1}
(m_i-m_c)^3 - \frac{\kappa_1}{(2J)^2}\,(J-J_c)\,(m_i-m_c) + \mathcal{O}\big((m_i-m_c)^4\big) = 0 \,,
\end{equation}
which, dividing by $m_i-m_c$ and solving, gives
\[ m_i-m_c = \pm\,\frac{\sqrt{\kappa_1}}{2J}\,(J-J_c)^\frac{1}{2} + \mathcal{O}\big((m_i-m_c)^\frac{3}{2}\big) \;;\]
hence $m_i-m_c\sim\sqrt{\kappa_1}/(2J)\,(J-J_c)^{1/2}$, proving the result (along the sequence).\\[4pt]
\textit{2)} Now suppose $(m_i-m_c)^4 = o\big(\gamma''(\bar J)\,(J-J_c)^2\big)$ (along a sequence), then (\ref{eq: critical equation for gamma}) rewrites
\begin{equation} \label{eq: critical equation for gamma 2}
(m_i-m_c)^3 \underbrace{- \frac{\kappa_1}{(2J)^2}\,(J-J_c)}_{=:\,p}\;(m_i-m_c) \underbrace{- \frac{\kappa_2}{(2J)^3}\,\gamma''(\bar J)\,(J-J_c)^2\,(1+o(1))}_{=:\,q} = 0 \,.\\[-5pt]
\end{equation}
Claim $\Delta:=(\frac{q}{2})^2+(\frac{p}{3})^3\leq0$.
Suppose by contradiction $\Delta>0$. Then the cubic equation (\ref{eq: critical equation for gamma 2}) has only one real solution: for $i=1,2$ %(see proposition \ref{propA: third degree equations})
\[ m_i-m_c = u_+ + u_- \quad\text{with }\, u_\pm = \sqrt[3]{-\frac{q}{2}\,\pm\sqrt[2]{\big(\frac{q}{2}\,\big)^2+\big(\frac{p}{3}\,\big)^3}} \ .\]
Observe that $q$ and $p$ are written only in terms of $J$, so that $u_+ + u_-$ at the main order do not depend implicitly on $m_i$.
%We may write explicitly
%\[ u_\pm =
%\sqrt[3]{-\frac{\kappa_2}{2(2J)^3}\,\gamma''(\bar J)\,(J-J_c)^2\,(1+o(1)) \pm
%\sqrt[2]{\frac{\kappa_2^2}{4(2J)^6}\,\gamma''(\bar J)^2\,(J-J_c)^4\,(1+o(1))-\frac{\kappa_1^3}{27(2J)^6}\,(J-J_c)^3}} \;,\]
%which, at the main order, do not depend on $m_i$.
Therefore $m_1-m_c$ and $m_2-m_c$ must have the same sign for every $J>J_c$ small enough. But this contradicts proposition \ref{prop: stationary points of vp} and lemma \ref{lem: flex exponent}, which ensures that in a right neighbourhood of $J_c$
\[ m_2-m_c > \phi_2-m_c > 0 \quad\text{while }\; m_1-m_c < \phi_2-m_c < 0 \;.\]
%On the other hand observe that, as $J>J_c$, $p<0$ and thus
%\[ \Delta>0\ \Rightarrow\ \big(\frac{p}{3}\,\big)^3=\mathcal{O}\big(\big(\frac{q}{2}\,\big)^2\big)\ \Rightarrow\ %\Delta=\mathcal{O}\big(\big(\frac{q}{2}\,\big)^2\big) \;.\]
%Therefore
%\[ \xi_i-\xi_c \,=\, \mathcal{O}\big(\big(\frac{q}{2}\,\big)^\frac{1}{3}\big) \,=\, \mathcal{O}\big(\gamma''(\bar J)^\frac{1}{2}\,(J-J_c)^\frac{2}{3}\big) + \mathcal{O}\big((\xi-\xi_c)^\frac{4}{3}\big) \]
This proves $\Delta\leq0$.
And now adapting to equation (\ref{eq: critical equation for gamma 2}) the step \textit{ii.} of the proof of corollary \ref{cor: bounds for critical exponents}, $\Delta\leq0$ entails (along the sequence)
\[ m-m_c \,=\, \mathcal{O}\big((J-J_c)^\frac{1}{2}\big) \;.\]
This completes the proof of the proposition.
\endproof

%\begin{figure}[h]
%\centering
%\includegraphics[scale=0.7]{fig5.1.eps}
%\caption{Plot of the global maximum point $m^*$ against $h\,,J$. We can see the critical point and ``the wall''. $m^*$, in its %region of differentiability, is the thermodynamic limit of the monomer density in the imitative model on the complete graph %(remark \ref{rk: m*=monomer density}).}
%\end{figure}

To conclude the proof of theorem \ref{thm: critical exponents}, the part \textit{iii)}, regarding the behaviour of $m^*$ at $(h_c,J_c)$ along the ``wall'' curve $\Gamma$, is a consequence of the previous proposition. Indeed
\[ m^*(\gamma(J)^+,J)=m_2(\gamma(J),J)\,,\quad m^*(\gamma(J)^-,J)=m_1(\gamma(J),J) \]
for all $J>J_c$, by proposition \ref{prop: wall existence uniqueness} and continuity of $m_1,\,m_2$.
%
%To sum up, we can complete the proof of the main theorem.
%As observed in remark \ref{rk: global max m*}, the global maximum point $m^*$ is expressed piecewise using the two local maximum points $\mu_1$, $\mu_2$ and inherits their continuity property outside $\Gamma$ and their regularity properties outside $\overline\Gamma$.
%Thus the first part of the theorem is proved by proposition \ref{prop: continuity of stationary points of vp}.\\
%The second part of the theorem, regarding the behaviour of $m^*$ at the critical point $(h_c,J_c)$ along curve of class $C^2$ is a consequence of proposition \ref{prop: critical exponents}.\\
%The last part, regarding the behaviour of $m^*$ at $(h_c,J_c)$ along the ``wall'' curve $\Gamma$ is a consequence of proposition \ref{prop: critical exponents along wall}, because
%\[ m^*(\gamma(J)^+,J)=m_2(\gamma(J),J)\,,\quad m^*(\gamma(J)^-,J)=m_1(\gamma(J),J) \]
%for all $J>J_c$, by proposition \ref{prop: wall existence uniqueness} and continuity of $m_1,\,m_2$.
\endproof
\section*{Appendix}
\subsection*{A. Properties of the function $g$}
\renewcommand{\theprop}{A\arabic{prop}}
\setcounter{prop}{0}
\renewcommand{\thecor}{A\arabic{cor}}
\setcounter{cor}{0}
\renewcommand{\thelem}{A\arabic{lem}}
\setcounter{lem}{0}
\renewcommand{\theequation}{A\arabic{equation}}
\setcounter{equation}{0}
We study the main properties of the function $g$ defined by (\ref{eq: g(h)}), which are often used in the paper.
Remind
\[ g(h) \,=\, \frac{1}{2}\,(\sqrt{e^{4h}+4\,e^{2h}}-e^{2h}) \quad\forall\,h\in\R \;.\]

%\begin{figure}[H]
%\centering
%\includegraphics[scale=0.4]{fig4.eps}
%\caption{Plot of the function $g$.}
%\end{figure}

Standard computations show that $g$ is analytic on $\R$, $0<g<1$, $\lim_{h\to-\infty}g(h)=0$, $\lim_{h\to\infty}g(h)=1$, $g$ is strictly increasing, $g$ is strictly convex on $]-\infty,\frac{\log(2\sqrt{2}-2)}{2}]$ and strictly concave on $[\frac{\log(2\sqrt{2}-2)}{2},\infty[\,$, $g(\frac{\log(2\sqrt{2}-2)}{2})=2-\sqrt{2}$.

Solving in $h$ the equation $g(h)=k$ for any fixed $k\in\,]0,1[\,$, one finds the inverse function:
\begin{equation} \label{eq: inverse g(h)}
g^{-1}(k) \,=\, \frac{1}{2}\log\frac{k^2}{1-k} \quad\forall k\in\,]0,1[ \;.
\end{equation}
It is useful to write the derivatives of $g$ in terms of lower order derivatives of $g$ itself.
For the first derivative, think $g$ as $(g^{-1})^{-1}$ and exploit (\ref{eq: inverse g(h)}):
\begin{equation} \label{eq: g' function of g}
g'(h) \,=\, \frac{1}{(g^{-1})'(k)}_{\big|k=g(h)} =\, \frac{2\,k\,(1-k)}{2-k}_{\big|k=g(h)} =\, \frac{2\,g(h)\,(1-g(h))}{2-g(h)}
\end{equation}
Then for the second derivative, differentiate the rhs of (\ref{eq: g' function of g}) and substitute (\ref{eq: g' function of g}) itself in the expression:
\begin{equation} \label{eq: g'' function of g}
g'' =\, \frac{2\,g'}{2-g} \big(1-2\,g + \frac{g\,(1-g)}{2-g}\big) \,=\,
\frac{2\,g'\,(1-2\,g)+(g')^2}{2-g} \;.
\end{equation}
The same for the third derivative: differentiate the rhs of (\ref{eq: g'' function of g}) and substitute (\ref{eq: g'' function of g}) itself in the expression:
\begin{equation} \label{eq: g''' function of g} \begin{split}
g''' &=\, \frac{1}{2-g}\, \big(2\,g''(1-2\,g+g') -4\,(g')^2 + g'\,\frac{2\,g'\,(1-2\,g)+(g')^2}{2-g}\,\big) \,=\\
&=\, \frac{g''\,(2-4\,g+3\,g')-4\,(g')^2}{2-g} \;.
\end{split} \end{equation}

\begin{lem} \label{lemA: g'<c}
For $c > 6-4\sqrt{2}\,$,
\[ g'(\xi) < c \quad\forall\,\xi\in\R \;.\]
For $0<c\leq 6-4\sqrt{2}\,$,
\[ g'(\xi)
\,\begin{cases}
\,< c & \text{iff }\ \xi \,<\, \frac{1}{2}\,\log\alpha_-(c)\; \text{ or }\ \xi \,>\, \frac{1}{2}\,\log\alpha_+(c) \\[2pt]
\,> c & \text{iff }\ \frac{1}{2}\,\log\alpha_-(c) \,<\, \xi \,<\, \frac{1}{2}\,\log\alpha_+(c)
\end{cases} \;,\]
where
\[ \alpha_{\pm}(c) := \frac{-(c^2+8c-4)\,\pm\,(2-c)\sqrt{c^2-12c+4}}{4\,c} \ .\]
\end{lem}

\proof
Investigate for example the inequality $g'(\xi)< c$.
By (\ref{eq: g' function of g}) clearly $0<g'<2$, hence the inequality is trivially true for $c\geq 2$ and false for $c\leq 0$.\\
Using identity (\ref{eq: g' function of g}) one finds
\[ g'<c\ \ \Leftrightarrow\ \ 2\,g^2-(2+c)\,g+2c > 0 \;;\]
this is a second degree inequality in $g$ with $\Delta=c^2-12c+4$.\\
If $6-4\sqrt{2}<c<6+4\sqrt{2}$, it is verified for any value of $g$.\\
If instead $c\leq 6-4\sqrt{2}$ or $c\geq 6+4\sqrt{2}$, it is verified if and only if
\[ g(\xi) < \frac{2+c-\sqrt{c^2-12c+4}}{4}=:s_{-}(c) \quad\text{or}\quad g(\xi) > \frac{2+c+\sqrt{c^2-12c+4}}{4}=:s_{+}(c) \;.\]
For $0<c<2$, $\;s_\pm(c)\in\;]0,1[\,$ hence one can apply $g^{-1}$, which is strictly increasing:
\[ \xi < g^{-1}(s_{-}(c)) \quad\text{or}\quad \xi > g^{-1}(s_{+}(c)) \;.\]
This concludes the proof because identity (\ref{eq: inverse g(h)}) and standard computations show that
\[ g^{-1}(s_\pm(c)) \,=\, \frac{1}{2}\,\log\alpha_\pm(c) \;.\qedhere\]
\endproof
\subsection*{B. Technical results about implicit functions}
\renewcommand{\theprop}{B\arabic{prop}}
\setcounter{prop}{0}
\renewcommand{\thecor}{B\arabic{cor}}
\setcounter{cor}{0}
\renewcommand{\thelem}{B\arabic{lem}}
\setcounter{lem}{0}
\renewcommand{\theequation}{B\arabic{equation}}
\setcounter{equation}{0}
We report some useful technical results,  omitting the proofs.

The following proposition is a particular case of Berge's maximum theorem.

\begin{prop} \label{propA: continuity of the max, argmax}
Let $f\!:[0,1]\times\R^n\rightarrow\R$ and $c\!:\R^m\rightarrow[0,1]$ be continuous functions.
\begin{itemize}
\item[i.] The following function is continuous:
\[ F:\R^n\times\R^m\rightarrow\R\,,\quad F(x,y)=\max_{t\in[0,\,c(y)]}f(t,x) \]
\item[ii.] Suppose that for all $x,y\in\R^n$ the function $t\mapsto f(t,x)$ achieves its maximum on $[0,\,c(y)]$ in a unique point. Then also the following function is continuous:
\[ T:\R^n\times\R^m\rightarrow[0,1]\,,\quad T(x,y)=\,\argmax_{t\in[0,\,c(y)]}f(t,x) \]
\end{itemize}
\end{prop}

The next proposition is a partial statement of Dini's implicit function theorem. Then we give two simple corollaries which are used in the paper.

\begin{prop} \label{propA: Dini}
Let $F:\R^n\times\R\rightarrow\R$ be a $C^\infty$ function.
Let $(x_0,y_0)\in\R^n\times\R$ such that
\[ F(x_0,y_0)=0\,,\quad \frac{\partial F}{\partial y}\,(x_0,y_0)\neq0 \;.\]
Then there exist $\delta>0$, $\epsilon>0$ and a $C^\infty$ function $f\!:B(x_0,\delta)\rightarrow B(y_0,\epsilon)$ such that for all $(x,y)\in B(x_0,\delta)\times B(y_0,\epsilon)$
\[ F(x,y)=0\ \Leftrightarrow\ y=f(x) \,.\]
\end{prop}

\begin{cor} \label{corA: regularity of a continuous solution}
Let $F:\R^n\times\R\rightarrow\R$ be a $C^\infty$ function.
Let $\varphi:\R^n\rightarrow\R$ be a $\mathrm{continuous}$ function such that for all $x\in\R^n$
\[ F\big(x,\varphi(x)\big)=0\,,\quad \frac{\partial F}{\partial y}\,\big(x,\varphi(x)\big)\neq0 \;.\]
Then $\varphi\in C^\infty(\R^n)$.
\end{cor}

\begin{comment}
\proof
Let $x_0\in\R^n$ and $y_0=\varphi(x_0)$. Let $f\!:B(x_0,\delta)\rightarrow B(\varphi(x_0),\epsilon)$ be the $C^\infty$ function of proposition \ref{propA: Dini}.\\
Since $\varphi$ is continuous, there exists $0<\delta'\leq\delta$ such that $\varphi\big(B(x_0,\delta')\big)\subseteq B\big(\varphi(x_0),\epsilon\big)$.
Therefore, by proposition \ref{propA: Dini},  for all $x\in B(x_0,\delta')$
\[ F\big(x,\varphi(x)\big)=0\,,\ (x,\varphi(x))\in B(x_0,\delta)\times B(\varphi(x_0),\epsilon)\ \Rightarrow\ \varphi(x)=f(x) \,.\]
Thus $\varphi\in C^\infty(B(x_0,\delta'))$ and this is sufficient by arbitrariness of $x_0$.
\endproof
\end{comment}

\begin{cor} \label{corA: regularity of a unique solution}
Let $F:\R^n\times\R\rightarrow\R$ be a $C^\infty$ function. Let $a,b:\R^n\rightarrow\R$ be continuous functions, $a<b$.
Suppose that for all $x\in\R^n$ there exists a $\mathrm{unique}$ $y=\varphi(x)\in\;]a(x),b(x)[$ such that
\[ F\big(x,\varphi(x)\big)=0 \;.\]
Moreover suppose that for all $x\in\R^n$, $\dfrac{\partial F}{\partial y}\,\big(x,\varphi(x)\big)\neq0\,$.
Then $\varphi\in C^\infty(\R^n)$.
\end{cor}

\end{document}